\documentclass[
reprint,
superscriptaddress,
amsmath,amssymb,
aps,
pra,
showkeys,
]{revtex4-2}

\usepackage{graphicx}
\usepackage{dcolumn}
\usepackage{bm}
\usepackage{hyperref}
\usepackage{braket}
\usepackage{appendix}
\usepackage{lipsum}

\usepackage{xr}

\begin{document}
\title{Shared control of a 16 semiconductor quantum dot crossbar array}

\author{Francesco Borsoi}
\author{Nico W. Hendrickx }
\author{Valentin John}
\author{Sayr Motz}
\author{Floor van Riggelen}
\affiliation{QuTech and Kavli Institute of Nanoscience, Delft University of Technology, P.O. Box 5046, 2600 GA Delft, The Netherlands}
\author{Amir Sammak}
\affiliation{QuTech and Netherlands Organisation for Applied Scientific Research (TNO), 2628 CK Delft, The Netherlands}
\author{Sander L. de Snoo}
\author{Giordano Scappucci}
\author{Menno Veldhorst}
\affiliation{QuTech and Kavli Institute of Nanoscience, Delft University of Technology, P.O. Box 5046, 2600 GA Delft, The Netherlands}
	
\begin{abstract}
	The efficient control of a large number of qubits is one of most challenging aspects for practical quantum computing. 
	Current approaches in solid-state quantum technology are based on brute-force methods, where each and every qubit requires at least one unique control line, an approach that will become unsustainable when scaling to the required millions of qubits. 
	Here, inspired by random access architectures in classical electronics, we introduce the shared control of semiconductor quantum dots to efficiently operate a two-dimensional crossbar array in planar germanium. 
	We tune the entire array, comprising 16 quantum dots, to the few-hole regime and, to isolate an unpaired spin per dot, we confine an odd number of holes in each site. 
	Moving forward, we establish a method for the selective control of the quantum dots interdot coupling and achieve a tunnel coupling tunability over more than 10 GHz. 
	The operation of a quantum electronic device with fewer control terminals than tunable experimental parameters represents a compelling step forward in the construction of scalable quantum technology. 
\end{abstract}

\maketitle

Fault-tolerant quantum computers will require millions of interacting qubits~\cite{Fowler2012,Wecker2014, Terhal2015QuantumMemories}. 
Scaling to such extreme numbers imposes stringent conditions on all the hardware and software components, including their integration~\cite{Vanmeter2013}. 
In semiconductor technology, decades of advancements have led to the integration of billions of transistor components on a single chip. 
A key enabler has been the ability to control such a large number of components with only a few hundred to a few thousand control lines~\cite{Landman1971,Vandersypen2017InterfacingCoherent}. 
In quantum technology, such a game-changing strategy has yet to be embraced owing to the fact that qubits are not sufficiently similar to each other.
Nowadays, leading efforts in solid state, such as superconducting and semiconducting qubits, all require that each and every qubit component is connected to at least one unique control line~\cite{Franke2019RentsComputing}. 
Clearly, this brute-force approach is not sustainable for attaining practical quantum computation.\\	
The development of spin qubits in semiconductor quantum dots has strongly been inspired by classical semiconductor technology~\cite{Loss1998QuantumDots, Zwanenburg2013SiliconElectronics, Chatterjee2021SemiconductorPractice}. 
Advanced semiconductor qubit systems are based on CMOS compatible materials and even foundry manufactured qubits have been realised~\cite{Maurand2016, Zwerver2022}.
In addition, it is anticipated that the small qubit footprint and compatibility with (cryo-)CMOS electronics will open up avenues to build integrated quantum circuits~\cite{Veldhorst2017SiliconComputer, Xue2021}. 
To enable the efficient control of larger qubit architectures with a sustainable number of control lines, proposals of architectures inspired by classical random access systems have been put forward~\cite{Hill2015QuantumSilicon,Li2018}.
However, their practical realisation has been so far prevented by device quality and material uniformity.\\
\begin{figure*}[htp!]
	\centering
	\includegraphics{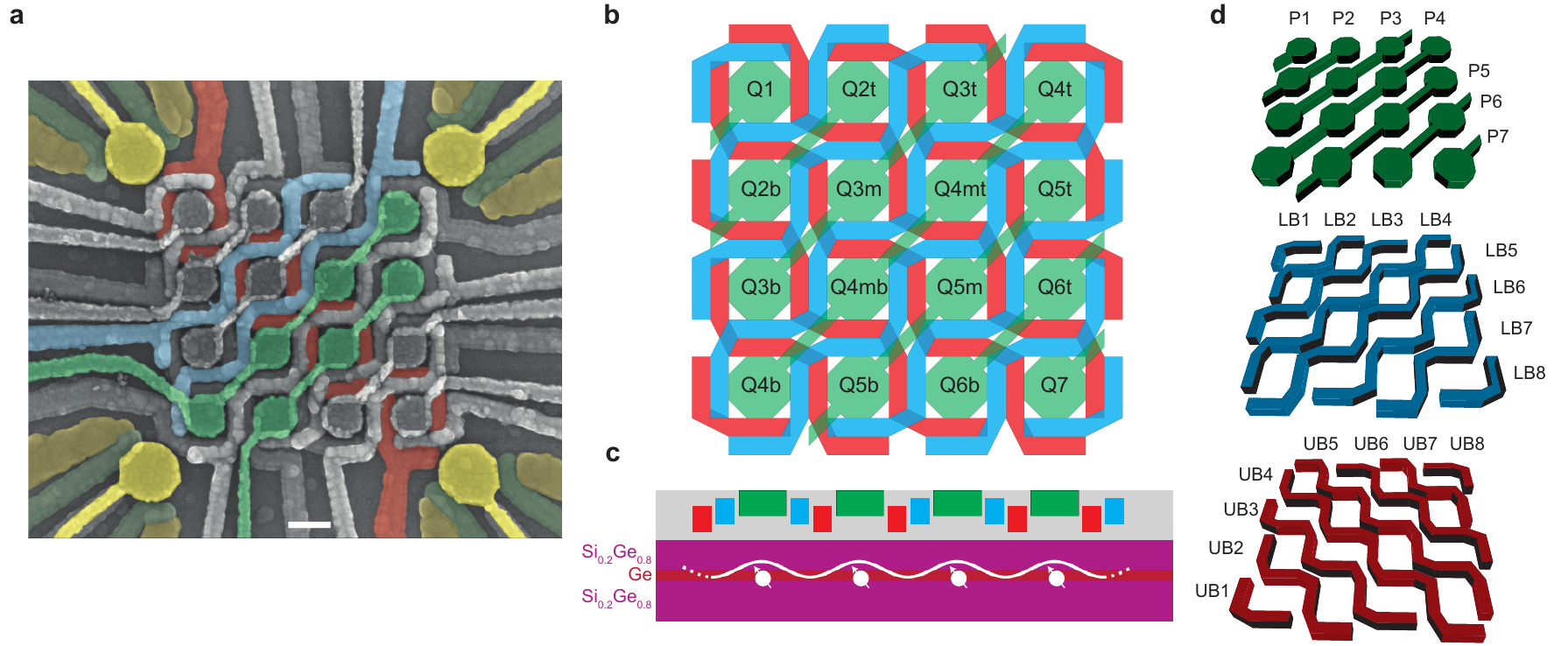}
	\caption{\textbf{A 16 quantum dot crossbar array.} 
		\textbf{a,}~False-colour scanning electron microscopy image of the crossbar array device. 
		The architecture consists of two staircase barrier gate layers (two lines of each layer are shown in red and blue), and one plunger gate layer (two lines are illustrated in green). 
		16 quantum dots are defined under the plungers, while four charge sensors in the form of single-hole transistors are located at the corners (ohmic contacts in orange, barriers in dark green, and plungers in yellow). 
		The scale bar corresponds to 100 nm, which is also the size of the designed plunger gates diameter. 
		This shared-control approach enables to control a number of quantum dots ($g$) with a sublinear number of control terminals ($T$). Here, the scaling is given by  $T = 6 g^{1/2} - 1$ (Suppl. Fig.~1).
		\textbf{b,}~Schematic illustration of the device and labelling of each quantum dot. 
		We choose to label the quantum dots after their positions on their controlling plunger line, e.g., the quantum dot Q6b(t) is located on the bottom (top) site controlled by the plunger line P6.
		\textbf{c,}~Schematic cross-section of the device. Holes are isolated in quantum dots in a 55 nm-deep germanium quantum well in a silicon-germanium heterostructure grown on a silicon wafer. 
		\textbf{d,}~Shared-control elements: from the bottom of the gate stack, eight UB barrier gates, eight LB barrier gates and seven P plunger gates. The overlay of these layers is visible in \textbf{(b)}.}
	\label{fig:device}
\end{figure*}
Here, we take the first step toward the sustainable control of large quantum processors by operating semiconductor quantum dots in a crossbar architecture.
This strategy enables the manipulation of the most extensive semiconductor quantum device with only a few shared-control terminals.
This is accomplished by exploiting the high quality and uniformity of strained germanium quantum wells~\cite{Scappucci2021}, by introducing an elegant gate layout based on diagonal plunger lines and double barrier gates, and by establishing a method that directly maps the transitions lines of charge stability diagrams to the associated quantum dots in the grid. 
We operate a two-dimensional 16 quantum dot system and demonstrate the tune-up of the full device to the few-hole regime.
In this configuration, we also prove the ability to prepare all the quantum dots in the odd charge occupation, as a key step for the confinement of an unpaired spin in each site~\cite{Hanson2007, Lawrie2020Benchmarks}. 
We then introduce a random access method for addressing the interdot tunnel coupling and find a remarkable agreement in the response of two vertically- and horizontally-coupled quantum dot pairs. 
We also discuss some critical challenges to efficiently operating future large quantum circuits.
\begin{figure}[htp]
	\centering
	\includegraphics{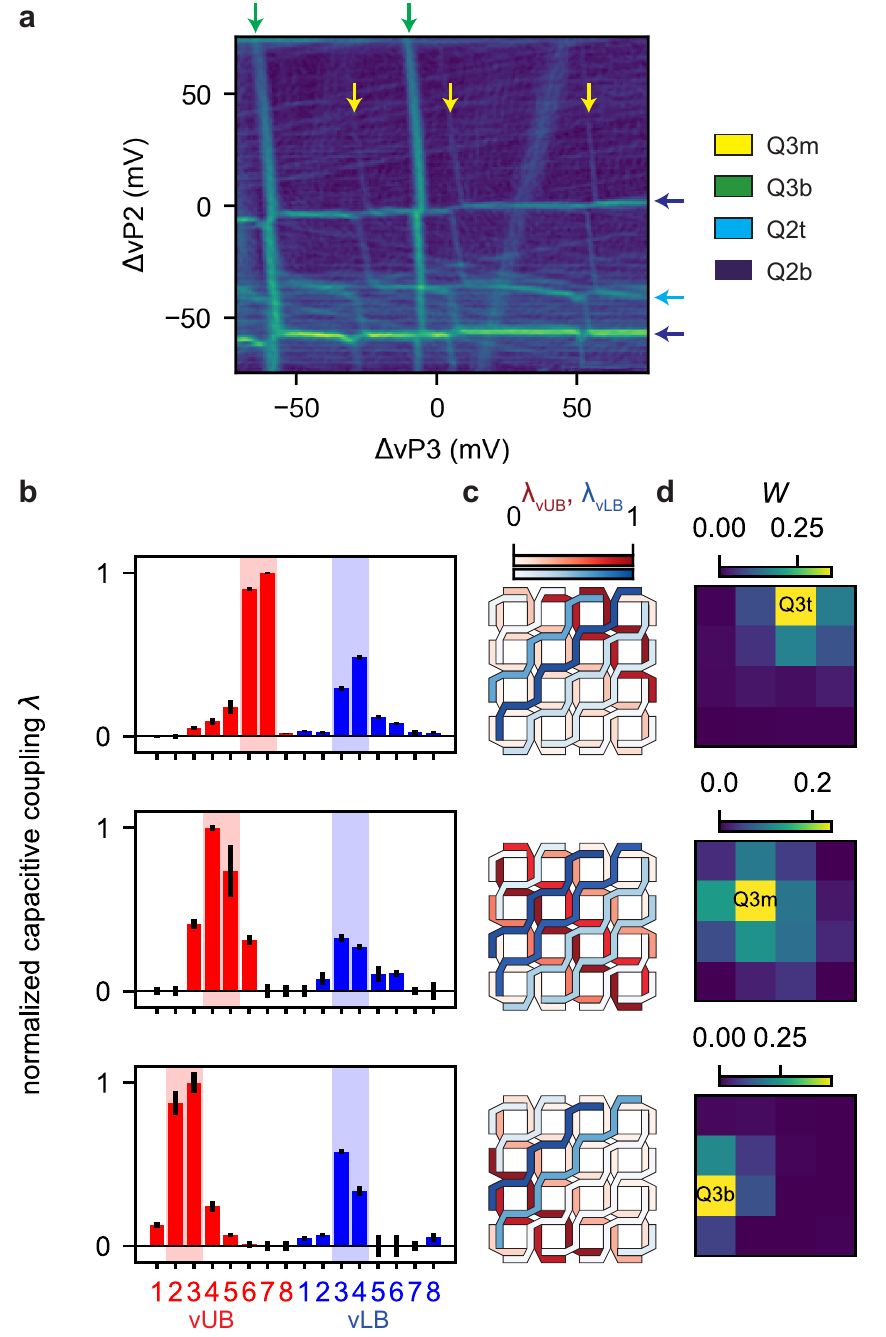}  
	\caption{\textbf{From multi-dot charge stability diagrams to quantum dots identification.}
		\textbf{a,} Few-hole charge stability diagram obtained by combining the signal gradients of the SW and NW charge sensors. 
		The labels indicate for each addition line the corresponding quantum dot.
		At (0,0) the fillings of Q2t, Q2b, Q3t, Q3m, Q3t are 0, 0, 0, 3, 0, respectively. 
		\textbf{b,} Histograms of the (normalized) capacitive couplings $\lambda$ of the barrier gates to the transition lines (in red for the vUB, and in blue for the vLB virtual gates) obtained by analysing three different sets of transition lines coupled to the virtual plunger vP3. Middle and bottom panels are extracted from the shift of the transition lines at $\sim$ (-20, -63) and (-15, -7) in \textbf{a}, respectively.
		The top panel (relative to dot Q3t) is obtained from a measurement shown in Suppl. Figs.~6g,h, where the transition line is more visible.
		We find that due to the order in the gate stack, the capacitive coupling of the vUB gates is a factor of 2-3 higher than of the vLB gates.
		Red and blue backgrounds are added to emphasize the two barriers with the highest couplings. 
		\textbf{c,} Device layout with the capacitive couplings colour-coded on the filling of the gate lines. Here, both the vUB and vLB capacitive couplings, $\lambda_{\mathrm{vUB}}$ and $\lambda_{\mathrm{vLB}}$, are normalized to their maximum values.
		At an intuitive level, the quantum dot associated with the analysed transition lines is located at the intersection of the two intensely coloured red and blue lines.
		\textbf{d,} Visualisation of the probability ($W$) calculated from of the shift of the three sets of addition lines (see Methods for details). 
		The comparison of the top, middle and bottom panels of \textbf{b, c, d} clearly distinguishes the three Q3 quantum dots (full analysis in Suppl. Figs.~4 and 5).
	}
	\label{fig:charge_sensing}
\end{figure}
\section{A two-dimensional quantum dot crossbar array} 
Our shared-terminals control approach for a two-dimensional quantum dot array with dots Q1, Q2t,...Q7 is based on a multi-layer gate architecture (Figs.~\ref{fig:device}a-d). 
We use two barrier layers (with gates UB$i$ and LB$i$ with $i \in [1, 8]$) to control the interdot tunnel couplings, and exploit a layer of plunger lines (P$i$ with $i \in [1,7]$) to vary the on-site energies (Fig.~\ref{fig:device}d).
In contrast to brute-force implementations, a single plunger gate is here employed to control up to four quantum dots, and an individual barrier up to six nearest-neighbours interactions.
In analogy with classical integrated circuits, this strategy enables to manage a number of experimental parameters (i.e., dot energies and interdot couplings) with a sub-linear number of control terminals, an approach that may overcome, among other aspects, the wiring interconnect bottleneck of large-scale spin qubit arrays (Suppl. Fig.~1)~\cite{Vandersypen2017InterfacingCoherent, Franke2019RentsComputing}.\\
To monitor the quantum dot array, we make use of charge sensing techniques~\cite{Hanson2007}. 
Four single-hole transistors at the corners of the array (named after their cardinal positions: NW as north-west, NE as north-east, SW as south-west and SE as south-east) act as charge sensors as well as holes reservoirs for the array.
The simultaneous read-out of their electrical response in combination with fast rastering pulse schemes enables us to continuously measure two-dimensional charge stability diagrams in real time (Methods)~\cite{Hendrickx2019_2Qubits, Vigneau2022}. \\
To bring the device in the 16 quantum dots configuration, we identify an alternative strategy to the tune-up methods established for individually-controlled quantum dot devices~\cite{Volk2019LoadingRegister}.
We begin by lowering all gate voltages to starting values based on previous experiments (Methods).
We set up the four independent charge sensors (Suppl. Fig.~2) and start defining a set of virtual gates as linear combinations of real gates~\cite{Hensgens2017, Mills2019ShuttlingDots, Volk2019LoadingRegister}. 
Throughout our work, we refer to vP1,..., vP7, vUB1,..., vUB8, vLB1,..., vLB8 as the virtual gates associated to the actual gates P1,..., P7, UB1,..., UB8, LB1,..., LB8 (Methods and Suppl. Fig.~3).
By further adjusting the respective virtual plunger and barrier voltages, we then accumulate the first holes in the quantum dots at the four corners of the array.  
Once clear charge addition lines are detected in the charge stability diagrams, we proceed with the tune-up of the adjacent dots and finish with the quantum dots furthest to the sensor.  
Owing to the homogeneity of our heterostructure~\cite{Lodari2021} and the symmetry of our gate layout, the accumulation of the first few holes in quantum dots controlled by the same plungers occurs at similar gate voltages. \\
Rather, challenges in the tuning up the array that we encounter are mainly due to elements outside the array.
In fact, we observe that small variations of the gate voltages impact the electrostatics of the dense gate fanout area, which in turn affects the charge sensors and the readout quality.
Furthermore, specific gate settings cause unintentional quantum dots under the gate fanout, which restrict the operational window. 
Altogether, these issues are a challenge for the implementation of automated tuning methods~\cite{Dawid2022}. 
We envision that the integration of a lower layer of ``screening" gates or the implementation of the gate fanout in the third dimension~\cite{Ha2022} can mitigate this issue (Suppl. Note 4).
\section{From multi-dot charge stability diagrams to quantum dots identification}
Moving forward, a direct result of our control approach is the fact that, upon sweeping the voltage of a plunger gate controlling $n$ quantum dots, up to $n$ sets of charge transitions can be observed, each associated with (un)loading an additional hole in one of the $n$ quantum dots. 
For the case of the two shared virtual plungers vP2 and vP3, this results in the charge stability diagram shown in Fig.~\ref{fig:charge_sensing}a where a number of vertical and horizontal charge addition lines marking well separate charge states are visible. 
However, because of our control approach, a priori it is unknown to which of the Q3 (Q2) quantum dots these vertical (horizontal) lines are associated. \\
Here, we solve this problem by establishing a statistical method that maps such transition lines to the respective quantum dot. 
In our protocol, we first evaluate the shift of the charge transition lines induced by a voltage variation of each barrier gate to estimate the (normalized) capacitive coupling $\lambda$ between each barrier gate and the associated quantum dot (Figs.~\ref{fig:charge_sensing}b, c and Methods). 
Because the two barrier layers form a grid of lines and columns crossing the device, we then use their capacitive couplings to infer the spatial location of the quantum dot in the array.
For this purpose, we consider the normalised capacitive couplings of the two orthogonal barrier sets, $\lambda_{\mathrm{vUB}}$ and $\lambda_{\mathrm{vLB}}$, as two independent probability distributions.
We then use $\lambda_{\mathrm{vUB}}$ and $\lambda_{\mathrm{vLB}}$ to calculate the combined probability $W$ on each of the 16 sites (Fig.~\ref{fig:charge_sensing}d, Methods). 
Finally, our protocol ends assigning the site with the maximum probability to the quantum dot loaded via the specific charge addition lines. 
In practice, $W$ quantifies how much an electric field generated on each site is perceived by a hole in a specific quantum dot site.
Hence, a low (high) $W$ value identifies a location that is weakly (strongly) coupled to the analysed quantum dot. \\
In Figs.~\ref{fig:charge_sensing}b-d, we show how this routine is effective for distinguishing and characterising the three Q3 quantum dots, whereas similar results are obtained also for the remaining dots in the grid (Suppl. Figs.~4 and~5). 
The demonstrated ability of labelling multi-quantum dot charge stability diagrams makes this method an important tool in the tune-up of large quantum dot devices.
We note that we can obtain further confirmation of our identification by also considering other aspects, such as the different sensitivity of the charge sensors to the same set of charge addition lines (e.g. compare Suppl. Figs.~6e, j, c). 
\section{Quantum dot occupancies} 
\begin{figure}[htp]
	\centering
	\includegraphics{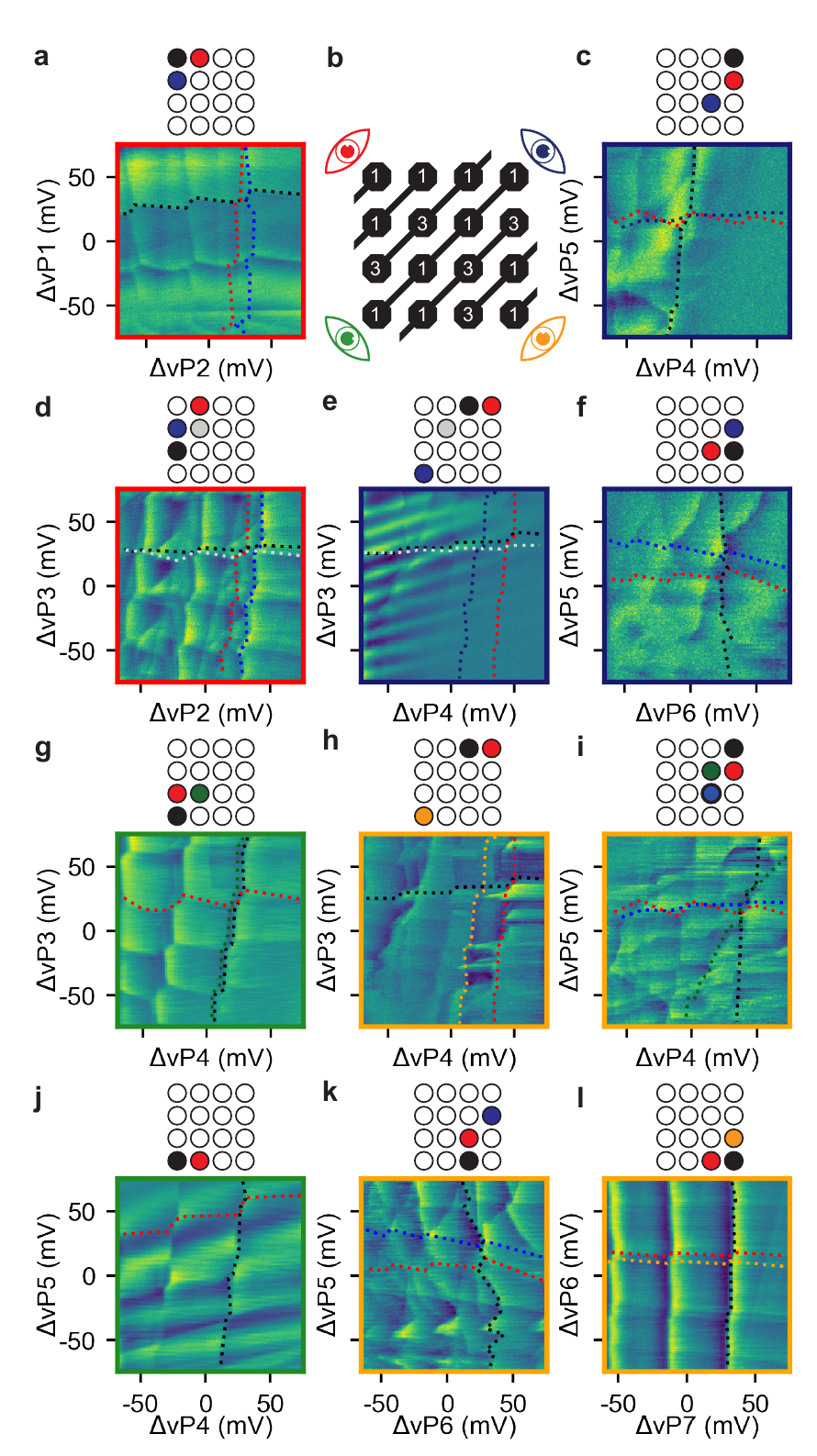}
	\caption{\textbf{Few and odd charge occupancy in a crossbar array.}
		\textbf{a, c-l,}~Charge stability diagrams displaying the sensor signals (after subtraction of a slowly-varying background) showcasing the 16 quantum dot system.
		At the centre of all the maps ($\Delta$vP$i$ = 0 with $i \in [1, 7]$), all the quantum dots are set in a odd occupancy regime.
		In each map, the first visible transition lines from the right and the top are labelled and assigned to the relative quantum dot by dashed lines (as guides for the eyes) with colours defined as in the schematic above each panel. 
		The colour of the panel frame identifies the sensor used: NW in red, NE in blue, SW in green, and SE in ochre. 
		Preparing the system in such a configuration results in the occupation of unintentional quantum dots under the fanout, which mostly do not interact with the dots inside the array 
		(e.g. see horizontal transition in (\textbf{a}) at $\Delta$vP1 = -50 mV and the periodical pattern in \textbf{(e)}). 
		\textbf{b,}~Schematic with the shared plungers layout together with quantum dot hole fillings, as monitored by the four charge sensors depicted as four eyes. }
	\label{fig:crossbar_odd}
\end{figure}

Whilst useful in reducing drastically the number of control terminals, a crossbar approach is effective for spin-based quantum computing if it enables to isolate a single or an unpaired spin in the individual quantum dots~\cite{Loss1998QuantumDots, Li2018}. 
Here, by harnessing the low disorder of the heterostructure and the symmetry of our design, we demonstrate the tune-up of the array to an odd charge occupancy configuration with 11 quantum dots filled with one hole, and five quantum dots filled with three holes (Fig.~\ref{fig:crossbar_odd}b).
Figs.~\ref{fig:crossbar_odd}a,c-l present a set of charge stability diagrams where the transition lines of each quantum dots are shown and labelled at least once. 
Importantly, at the centre of the panels (i.e., $\Delta$vP$i$ = 0, with $i \in [1, 7]$), each quantum dot is filled with an odd number of holes, hence with an unpaired hole spin (Methods).
We note that holes in quantum dots that are located in the core of the array are loaded/unloaded by means of cotunneling processes via the outer dots~\cite{DeFranceschi2001}. 
When such reservoir-dot tunnelling timescale becomes long and approaches the timescale of our measurement scan ($ \sim 0.01$ s), the quantum dots transition lines become weakly visible in the charge stability diagrams and appear distorted due to latching effects, explaining some of the features observed in Figs.~\ref{fig:crossbar_odd}h, i, k.~\cite{Eenink2019,Lawrie2020}. 
Additional reservoirs within the array may therefore simply the tune-up, though for quantum operation is it not critical to have each and every dot strongly coupled to a reservoir~\cite{Vandersypen2017InterfacingCoherent, Chatterjee2021SemiconductorPractice}. \\
\begin{figure*}[htp]
	\centering
	\includegraphics{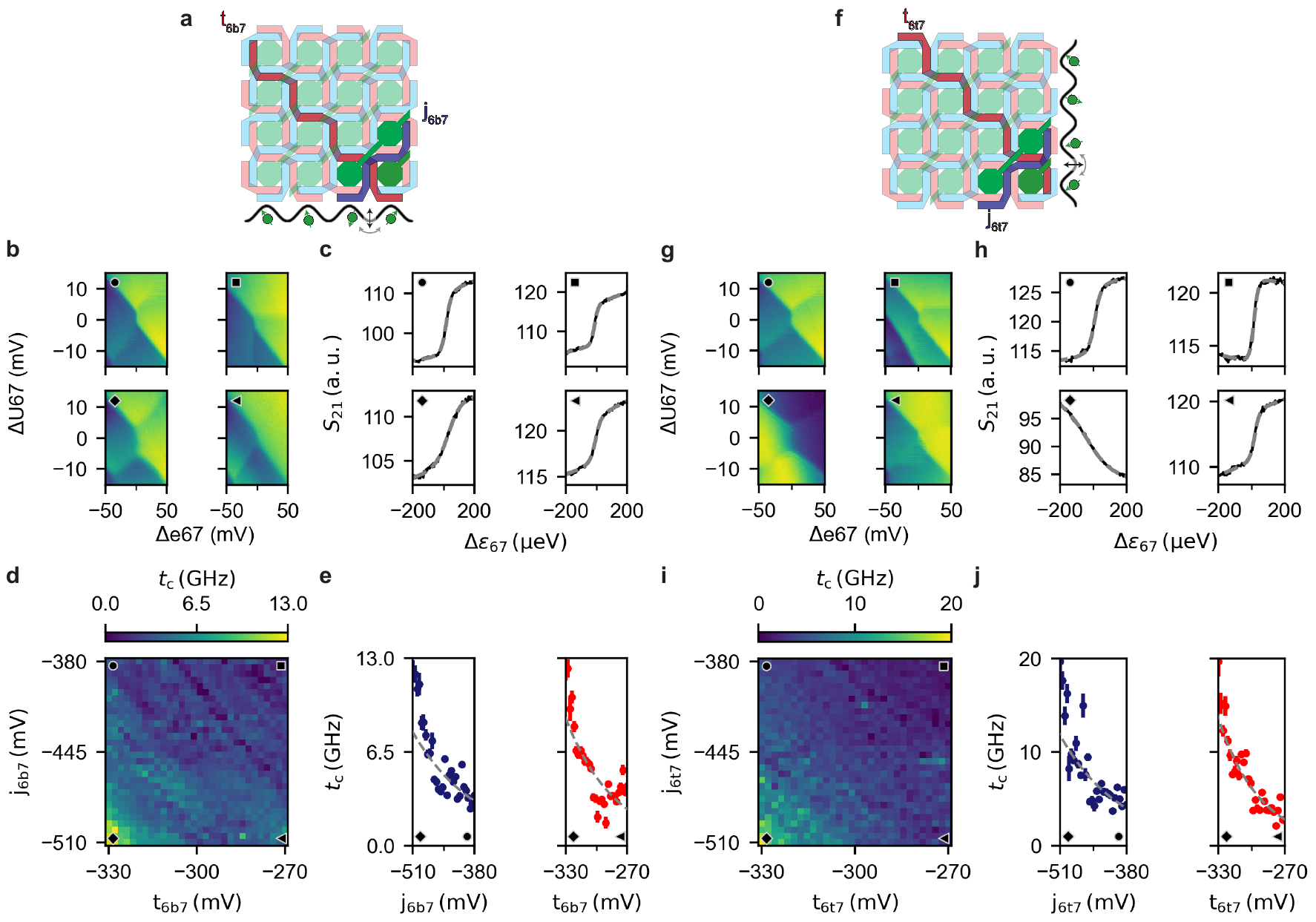}
	\caption{\textbf{Addressable control of the interdot tunnel coupling using double barrier gates.}
		\textbf{a, f} Schematics of the crossbar indicating the two intersecting barriers (in red and blue) controlling the Q6b-Q7 and Q6t-Q7 interdot tunnel couplings, respectively. 
		\textbf{b, g} Exemplary charge stability diagrams taken via reflectometry methods showing the Q6b-Q67 and Q6t-Q7 (3,1)-(2,2) charge states at four different virtual barrier voltages configurations, respectively. 
		At the centre of the panels, the vertical interdot transition line is clearly visible.
		The circle, square, diamond, and triangle markers correlate each map to the voltages, $(\mathrm{t_{6b7}}, \mathrm{j_{6b7}}) = (\mathrm{t_{6t7}}, \mathrm{j_{6t7}}) = -(330, 380), (270, 380), (330, 510), (270, 510) $ mV, respectively.  
		\textbf{c, h}  Charge polarisation traces  (black) relative to panels \textbf{(b, g)}, together with best fit (dashed grey). 
		\textbf{d, i} Colour maps of the two-axis controlled tunnel couplings of the systems Q6b-Q7 and Q6t-Q7, with markers located at the respective voltages. 
		\textbf{e, j} Vertical (left panel) and horizontal (right panel) linecuts of \textbf{(d)} and \textbf{(i)}  at $\mathrm{t_{6b7}} = \mathrm{t_{6t7}}  = -330$ mV and $\mathrm{j_{6b7}} = \mathrm{j_{6t7}} =-510$ mV, respectively. 
		Grey traces are fits with an exponential function of the data, from which we obtain the four effective barrier lever arms (Methods).
	}
	\label{fig:tunnel_coupling}
\end{figure*}
\section{Interdot couplings control}
The ability to selectively tune the interdot coupling in a quantum dot architecture is crucial for generating exchange-based entanglement between semiconductor qubits~\cite{Loss1998QuantumDots}.
Here, inspired by the word and bitlines approach as in dynamic random-access memories~\cite{Veldhorst2017SiliconComputer, Li2018}, we exploit the double barrier design to spatially define and activate unique points in the grid structure.
Conceptually, each two-barrier intersection point can be set by the respective voltages in the four configurations: (ON, ON), (ON, OFF), (OFF, ON), (OFF, OFF). 
For selective two-qubit operations in qubit arrays, the voltage set points should be calibrated such that only when both barriers are in the ON-state a two-qubit interaction is activated, leaving all the other pairs non-interacting (Suppl. Fig.~7). \\
Here, we implement a proof-of-principle of this method by demonstrating the two-barrier control of the interdot tunnel coupling.
To this end, we investigate the tunnel coupling variations of the horizontal Q6b-Q7 and vertical Q6t-Q7 quantum dot pairs as a function of the two intersecting barriers.
Starting from the respective UB4 and LB7 gates, we define the virtual barriers $\mathrm{t_{6b7}}$ and $\mathrm{j_{6b7}}$ (Methods), which separate the quantum dots Q6b and Q7, while keeping their detuning (e67) and on-site (U67) voltages constant at the (3,1)-(2,2)  Q6b-Q7 interdot transition (Fig.~\ref{fig:tunnel_coupling}a). After obtaining the detuning lever arm $\alpha_{\epsilon_{\mathrm{67}}}$ to convert the detuning voltage into an energy scale ($\Delta \epsilon_{\mathrm{67}} = \alpha_{\epsilon_{\mathrm{67}}}  \cdot \Delta$e67), we evaluate the strength of the interdot couplings by analysing the charge polarisation lines along the detuning axis at $\Delta $U67 = 0 (Figs.~\ref{fig:tunnel_coupling}b, c)~\cite{DiCarlo2004DifferentialDot}.
By performing this measurement systematically, we demonstrate that the tunnel coupling can be controlled effectively by both barriers (Figs.~\ref{fig:tunnel_coupling}d,e). 
At values of $(\mathrm{t_{6b7}}, \mathrm{j_{6b7}}) = (-270, -380)$ mV, the coupling is virtually OFF, but our method is inherently not accurate for $t_c \leq 3$ GHz, because the thermal energy dominates the broadening of the polarisation line~\cite{DiCarlo2004DifferentialDot, Eenink2019}. In contrast, upon activating both barriers at $(\mathrm{t_{6b7}}, \mathrm{j_{6b7}}) = (-330, -510)$ mV, the tunnel coupling is turned on following approximately an exponential trend (Methods). Within the displayed voltage range, we can tune it well above 10 GHz, much higher than in the configuration in which only one barrier is activated. 
We perform the same experiment on the dots pair Q6t-Q7 by defining the virtual barriers $\mathrm{t_{6t7}}$ and $\mathrm{j_{6t7}}$ based on the gates UB5 and LB7, respectively (Fig.~\ref{fig:tunnel_coupling}f and Methods). 
Using the same barrier voltages window as for Q6b-Q7, we find that the coupling tunability of the pair Q6t-Q7 is comparable to the previous pair, with a virtually OFF state ( $\leq $3 GHz) at $(\mathrm{t_{6t7}},\mathrm{j_{6t7}}) = (-270, -380)$ mV, and with a ON state reaching 20 GHz at $(\mathrm{t_{6t7}}, \mathrm{j_{6t7}}) = (-330, -510)$ mV (Figs.~\ref{fig:tunnel_coupling}g-j). 
These results are corroborated by the two-barrier tunability of the interdot capacitive coupling for both double-dot systems (Suppl. Fig.~8) and by the tunability of another dot pair (Suppl. Fig.~9).
We envision that for rapid qubits exchange operations at the charge symmetry point, the required barrier voltage window might be different from our measurable window via polarisation lines. 
Specifically, for state-of-the-art values of ON (OFF) exchange interaction of 50 MHz (10 KHz)  and a typical charging energy of 1 mV, the required tunnel coupling is $ \sim 2$ (0.02) GHz, which can be better calibrated via qubit spectroscopy techniques~\cite{Hendrickx2019_2Qubits, Hendrickx2021AProcessor, Xue2022}. 
\section{\label{sec:conclusion}Conclusions}
While semiconductor technology has gained credibility as a platform for large-scale quantum computation, experimental work primarily focused on small and linear arrays, and it is currently unclear how to scale quantum dot qubits to large numbers. 
Here, by implementing a strategy that allows to address a large number of quantum dots with a small number of control lines, we have operated the most extensive two-dimensional quantum dot array so far.
With this approach, the number of gate layers is independent of the grid size, which greatly simplifies the nanofabrication of quantum dot arrays.
With the introduction of a double barrier paradigm and a statistical method for labelling multi-dot charge stability diagrams, we have demonstrated two critical requirements for quantum logic in shared-control architectures: the tunability of 16 interacting quantum dots into an odd charge state with an unpaired spin and a method for the selective control of the interdot tunnel coupling, which is crucial for the control of the exchange interaction.  
We stress that, while the functionality of our approach crucially depends on the quality and homogeneity of the material stack, it can be applied to simplify the interconnection of quantum dot systems hosted in other material platforms, such as silicon~\cite{Lawrie2020, Philips2022}, gallium arsenide~\cite{Mortemousque2021, Fedele2021} and graphene~\cite{Gachter2022}. \\
Our experiment has also served to pinpoint a limit of our first implementation: additional stray dots at the periphery of the crossbar have complicated the tune-up of the device, which they may be resolved by additional screening gates in the fanout.
Moreover, the array is also reaching a size where manual tuning operations become a bottleneck. 
A promising direction to obtain the efficient control of quantum dot arrays is autonomous operation based on machine learning~\cite{Dawid2022}.
We envision that future crossbar arrays may find applications in large and dense two-dimensional quantum processors or as registers that are coupled via long-range quantum links for networked computing.
\bibliography{bibliography}
\section{Methods}
\noindent \textbf{Fabrication}\\
The device is fabricated on a Ge/SiGe heterostructure where a 16-nm-thick germanium quantum well with a maximum hole mobility of $2.5 \cdot 10^5  \, \mathrm{cm^2  \, V^{-1}  \,  s^{-1}}$ is buried 55 nm below the semiconductor/oxide interface~\cite{Sammak2019ShallowTechnology, Lodari2021}. 
We design the quantum dot plunger gates with a diameter of 100 nm, and the barrier gates separating the quantum dots with a width of 30 nm.
The fabrication of the device follows these main steps. 30 nm-thick Pt ohmic contacts are patterned via electron-beam lithography, evaporated and diffused in the heterostructure following an etching step to remove the oxidised Si cap layer~\cite{Hendrickx2018Gate-controlledGermanium, Tosato2022}.
A three-layer gate stack is then fabricated by alternating the atomic layer deposition of a $\mathrm{Al_2 O_3}$ dielectric film (with thicknesses of 7, 5, 5 nm) and the evaporation of Ti/Pd metallic gates (with thicknesses 3/17, 3/27, 3/27 for each deposition, respectively). 
After dicing, a chip hosting a single crossbar array is then mounted and wire-bonded on a printed circuit board. 
Prior to the cool down in a dilution refrigerated, we tested in a 4 K helium bath two nominally identical crossbar devices following the screening procedure shown in ref.~\cite{Hendrickx2021AProcessor}. Both devices exhibited full gates and ohmic contacts functionality, and one of them was mounted in a dilution refrigerator. \\

\noindent \textbf{Experimental setup}\\
The experiment is performed in a Bluefors dilution refrigerator with a base temperature of 10 mK. 
From Coulomb peaks analysis, we extract an electron temperature of $138 \pm 9$ mK, which we use to estimate the detuning lever arm (Suppl. Figs.~10 and~11).
We utilise an in-house built battery-powered SPI rack \url{https://qtwork.tudelft.nl/~mtiggelman/spi-rack/chassis.html} to set direct-current (dc) voltages, while we use a Keysight M3202A arbitrary waveform generator (AWG) to apply alternating-current (ac) rastering pulses via coaxial lines.
The dc and ac voltage signals are combined on the PCB with bias-tees and applied to the gates.
Each charge sensor is galvanically connected to a NbTiN inductor with an inductance of a few $\mathrm{\mu H}$ forming a resonant tank circuit with resonance frequencies of $\sim 100$ MHz. 
In our experiment we have observed only three out of four resonances, likely due to a defect inductor. 
Moreover, because two resonances overlap significantly, we mostly avoid using reflectometry (unless explicitly stated in the text) and use fast dc measurements with bandwidth up to 50 kHz.
The four dc sensor currents are converted into voltages, amplified and simultaneously read out by a four-channel Keysight M3102A digitizer module with 500 MSa/s. 
The digitizer module and several AWG modules are integrated into a Keysight M9019A peripheral component interconnect express extensions for instrumentation (PXIe) chassis. 
Charge stability diagrams here typically consists of a 150x150 pixels scan with a measurement time per pixel of 50 $\mathrm{\mu s}$. 
To enhance the signal-to-noise ratio, we average the same map 5-50 times, obtaining a high-quality map within a minute.
We adopt the notation $\Delta$g$i$ to identify an AWG-supplied ramp of the gate g$i$ with respect to a fixed dc reference voltage.\\

\noindent \textbf{Tune-up details}\\
The device was tuned into the 16-quantum-dots regime two times. 
In the first run, the gate voltages were optimised to minimise the number of unintentional quantum dots to better characterise the crossbar quantum dots (Fig.~\ref{fig:charge_sensing} and Suppl. Fig.~6). 
In the second, the stray dots were neglected and we focused on tuning the quantum dot array to the odd-occupation regime (Fig.~\ref{fig:crossbar_odd}). 
Between the two tune-up cycles, the gate voltages were reset to zero without thermal cycling the device.
The starting gate voltage values for the tune-up are -300 mV for the barriers and -600 mV for the plungers.
In Suppl. Fig.~12 we display the dc gate voltages relative to the measurements displayed in Fig.~\ref{fig:crossbar_odd}, with the crossbar array tuned in the odd-charge occupation. 
In this regime, we characterise the variability of the transition lines spacing to be $\sim 10-20\%$ as a metric for the level of homogeneity of the array (Suppl. Fig.~13)~\cite{Zajac2016}.
We could also have chosen the voltage onset of the first hole in each quantum dot, however these exact voltages depend on several surrounding gates, hence our choice on the transition lines spacing.
The odd occupancy is demonstrated by emptying each quantum dot as shown in Suppl. Figs.~14-25.
All the panels of Fig.~\ref{fig:crossbar_odd} are taken at the same gate voltage configuration on the same day, except for panels b, and d, which were retaken a few days later with improved sensitivity of the NW sensor and with minimal voltage differences (the largest variations are 12 mV on P7 which is far away from the Q1, Q2 and Q3 system and 11 mV on UB8). 
The original (vP2, vP1) and (vP2, vP3) measurements are shown in Suppl. Figs.~14 and~16.
During the experiment, the gate UB8 did not function properly, possibly due to a broken lead. To compensate for this effect and to enable charge loading in the dots P3t and P5t, we set UB7 at a lower voltage compared to the other UB gates.
Additionally, LB1 is set at a comparatively higher voltage to mitigate the formation of accidental quantum dots under the fanout of LB1 and P1 at lower voltages. The first addition line of such an accidental quantum dot is visible as a weakly interacting horizontal line in Fig.~\ref{fig:crossbar_odd}a. \\

\noindent \textbf{Virtual matrix}\\
We define a set of virtual gates ($\vec{\mathrm{v}G}$) as linear combinations of actual gates ($\vec{G}$) to eliminate the cross-talk to the charge sensors and to independently control the on-site energies with virtual plunger gates~\cite{Hensgens2017, Mills2019ShuttlingDots}. 
The matrix $M$ defined by $\vec{G} = M \cdot \vec{\mathrm{v}G}$  is shown as a colour map in Suppl. Fig.~3. 
For the tunnel coupling experiments presented in Fig.~\ref{fig:tunnel_coupling}, we employ additional virtual gate systems for achieving independent control of the detuning voltage e67 and U67, and of the interdot interactions via virtual barriers  $\mathrm{t_{6b7}}$, $\mathrm{j_{6b7}}$, $\mathrm{t_{6t7}}$ and $\mathrm{j_{6t7}}$. With SE\textunderscore P the SE plunger gate, we write:
{\begin{equation*}
		\begin{pmatrix}
			\text{P5} \\ \text{P6} \\ \text{P7}\\ \text{SE\textunderscore P} 
		\end{pmatrix}
		= 
		\begin{pmatrix} 
			0.04 & -1.2 \\
			-0.5 & 0.9 \\
			0.492 & 0.9 \\
			-0.08 & -0.26 \\
		\end{pmatrix}
		\begin{pmatrix} 
			\text{e67} \\
			\text{U67} \\
		\end{pmatrix}
\end{equation*}}
{\begin{equation*}
		\begin{pmatrix}
			\text{P6} \\ \text{P7} \\ \text{UB5}\\ \text{LB7} \\ \text{SE\textunderscore P}
		\end{pmatrix}
		= 
		\begin{pmatrix} 
			-1.28 & -0.33 \\
			-1.18 & -0.72 \\
			1 & 0\\
			0 & 1 \\
			0.15 & -0.01 \\
		\end{pmatrix}
		\begin{pmatrix} 
			\mathrm{t_{6t7}} \\
			\mathrm{j_{6t7}} \\
		\end{pmatrix}
\end{equation*}}
{\begin{equation*}
		\begin{pmatrix}
			\text{P6} \\ \text{P7} \\ \text{UB4}\\ \text{LB7} \\ \text{SE\textunderscore P}
		\end{pmatrix}
		= 
		\begin{pmatrix} 
			-2.05 & -0.97 \\
			-1.18 & -0.41 \\
			1 & 0\\
			0 & 1 \\
			-0.19 & -0.01 \\
		\end{pmatrix}
		\begin{pmatrix} 
			\mathrm{t_{6b7}} \\
			\mathrm{j_{6b7}} \\
		\end{pmatrix}
\end{equation*}}

\noindent \textbf{Quantum dots identification}\\
To obtain the capacitive coupling of all the barrier gates to a set of transition lines (as shown in Fig.~\ref{fig:charge_sensing}b), we acquire and analyse sets of 112 charge stability diagrams.
The same charge stability diagram is taken after stepping each barrier gate around its current voltage in steps of 1 mV in the range of -3 to 3 mV (i.e., 7 scans x 16 barriers).
In the analysis, we first subtract a slowly varying background to the data (with the function ndimage.gaussian.filter of the open-source scipy package) and then calculate the gradient of the map (with ndimage.gaussian\textunderscore gradient\textunderscore magnitude).
For a given line cut of such two-dimensional maps, we extract the peaks position using a Gaussian fit function.
Due to cross capacitance, the transition line positions manifest a linear dependence on each of the 16 barriers, which we quantify by extracting the linear slope (Suppl. Fig.~4). 
After normalisation to the maximum value, these parameters are named capacitive couplings ($\lambda$), and, because of the grid structure of the two barrier layers, provide a first information of where the hole is added/removed to/from.
To extract the quantum dot positions, we consider the capacitive couplings to the vUB ($\lambda_{\mathrm{vUB}}$) and vBL ($\lambda_{\mathrm{vLB}}$) gates as two independent probability distributions.
With this approach, the integral of the $\lambda_{\mathrm{vUB}}$ ($\lambda_{\mathrm{vLB}}$) between vUB$i$ (vLB$k$) and vUB$j$ (vLB$l$) returns a ``probability'' $p_{U, (i,j)}$ ($p_{L, (k,l)}$) to find the dot in-between these control lines. 
As a result, the combined probability in the site confined by these four barriers is given by the product of these elements: $w_{(i, j), (k, l)} = p_{U, (i,j)} \cdot p_{L, (k,l)}$.
We note that the sum of the 16 probabilities returns 1.  
As already observed in ref.~\cite{Lawrie2020}, the gates cross-coupling to a specific quantum dot defined in a germanium quantum well manifest a slow falloff in space (i.e., gates with a distance to the dot of $> 100$ nm still have a considerable cross-coupling to the dot).
This can be attributed to the rather large vertical distance between the gates and the quantum dots ($> 60$ nm), and is in contrast with experiments in SiMOS devices where the falloff is rather immediate due to the tight charge confinement.
This aspect explains why our probability W at the identified quantum dot reaches at a max of $0.25-0.5$.
\\~\\
\noindent \textbf{Tunnel coupling evaluation}\\
For the estimation of the tunnel coupling results presented in Fig.~\ref{fig:tunnel_coupling}, we established an automated measurement procedure that follows this sequence: 1) we step the virtual barriers across the two-dimensional map ($t, j$); 2) at each barrier configuration, we take a two-dimensional (e67, U67) charge stability map (Figs.~\ref{fig:tunnel_coupling}b-g); 3) we identify the accurate position of charge interdot via a fitting procedure of the map (Suppl. Fig.~8)~\cite{VanDiepen2018}; 4) we perform small adjustments at the e67, U67 virtual gates to centre the interdot at the (0,0) dc-offset; 5) measure the polarisation line by using $\sim$0.1 kHz AWG ramps (Figs.~\ref{fig:tunnel_coupling}c, h). For an accurate analysis, each polarisation line is the result of an average of 150 traces, using a measurement integration time of 50 $\mathrm{\mu} s$ per pixel. With this method, the full 30x30 maps are taken in a few hours. We fit the traces considering an electron temperature of 138 mK and a detuning lever arm of $\alpha_{\epsilon_{\mathrm{67}}} = 0.012(4)$ eV/V, extracted from a thermally broadened polarisation line (Suppl. Fig.~11). 
We observe that the extracted tunnel coupling follows approximately an exponential trend as a function of the barrier gates. We fit the data presented in Figs.~\ref{fig:tunnel_coupling}e, j with the function $A \cdot e^{-B V_g}$, with $A$ a prefactor, $B$ the effective barrier lever arm and $V_g$ the gate axis. We find that the effective barrier lever arms of $j_\mathrm{6b7}$ and $t_\mathrm{6b7}$ are  $0.007 \pm 0.002 $ and $0.021 \pm 0.003 \, \mathrm{mV^{-1}}$, respectively. Similarly, we find for $j_\mathrm{6t7}$ and $t_\mathrm{6t7}$ values of  $0.008 \pm 0.001 $ and $0.026 \pm 0.003 \, \mathrm{mV^{-1}}$, respectively. Altogether these results indicate that the lower barrier layer of UB gates is $\sim 3$ times more effective than the upper barrier layer of LB gates. This is consistent with what is found in Fig.~\ref{fig:charge_sensing}b and Suppl. Fig.~5. We note that for qubit operations in such crossbar array, it is actually necessary to fully characterise and calibrate the two-barrier tunability of all the 24 nearest-neighbours. Performing this task requires improving further our hardware implementation and is beyond the scope of this work. 
\section{Acknowledgements}
We are grateful to Corentin Depr\'{e}z, Chien-An Wang, Eli\v{s}ka Greplov\'{a} and all the members of the Veldhorst lab for fruitful discussions. 
We acknowledge support through an ERC Starting Grant and through an NWO projectruimte. Research was sponsored by the Army Research Office (ARO) and was accomplished under Grant No. W911NF-17-1-0274. The views and conclusions contained in this document are those of the authors and should not be interpreted as representing the official policies, either expressed or implied, of the Army Research Office (ARO), or the U.S. Government. The U.S. Government is authorized to reproduce and distribute reprints for Government purposes notwithstanding any copyright notation herein.
\section{Author contributions}
F.B., N.W.H. performed the experiments with the support of V.J. and F.v.R..
F.B. analysed the data and N.W.H fabricated the device.
V.J. and S.M. contributed to the data analysis.
S.L.d.S., N.W.H and F.B. developed the software used in the experiment.
A.S. and G.S. supplied the heterostructure.
F.B. and M.V. wrote the manuscript with the input from all the authors.
M.V. supervised the project.
\section{Competing interests}
The authors declare no competing interests.
Correspondence should be addressed to Francesco Borsoi (F.Borsoi@tudelft.nl).
\section{Data availability}
All data and analysis underlying this study are available on a Zenodo repository at \url{https://doi.org/10.5281/zenodo.7077895}~\cite{dataset}.
\end{document}


\title{Supplementary information: \linebreak Shared control of a 16 semiconductor quantum dot crossbar array}
\author{Francesco Borsoi}
\author{Nico W. Hendrickx }
\author{Valentin John}
\author{Sayr Motz}
\author{Floor van Riggelen}
\affiliation{QuTech and Kavli Institute of Nanoscience, Delft University of Technology, P.O. Box 5046, 2600 GA Delft, The Netherlands}
\author{Amir Sammak}
\affiliation{QuTech and Netherlands Organisation for Applied Scientific Research (TNO), 2628 CK Delft, The Netherlands}
\author{Sander L. de Snoo}
\author{Giordano Scappucci}
\author{Menno Veldhorst}
\affiliation{QuTech and Kavli Institute of Nanoscience, Delft University of Technology, P.O. Box 5046, 2600 GA Delft, The Netherlands}
\email{M.Veldhorst@tudelft.nl}
\maketitle

\renewcommand{\thepage}{S\arabic{page}} 
\renewcommand{\thesection}{Supplementary Note \arabic{section}}  

\setcounter{figure}{0}
\setcounter{page}{1}
\setcounter{section}{0}

\renewcommand{\figurename}{\textbf{Supplementary Figure}}
\renewcommand{\thefigure}{\textbf{\arabic{figure}}}

\renewcommand{\tablename}{\textbf{Supplementary Table}}
\renewcommand{\thetable}{\textbf{\arabic{table}}}

\renewcommand{\bibnumfmt}[1]{[S#1]}
\renewcommand{\citenumfont}[1]{S#1}
\onecolumngrid

\section{Rent's rule and gate count}
Scalable architectures impose stringent requirements at all layers of the quantum computing stack~\cite{Vanmeter2013}.
If we focus on the lowest layers of the computing stack, state-of-the-art solid-state quantum processors do not meet these prerequisites yet~\cite{Franke2019RentsComputing}.
In fact, current processors still make use of a few control terminals per qubit, an approach that will lead to arduous interconnectivity and control challenges in the route toward millions qubits~\cite{Arute2019, Philips2022}.
To quantify the level of optimisation and interconnectivity of a quantum processor, we borrow the concept of Rent's rule from classical electronics~\cite{Lanzerotti2005}. 
In quantum dot devices, the Rent's rule can be used to correlate the number of control terminals $T$ (i.e., gates, ohmic leads,...) and the number of active components $g$ (i.e., quantum dots or qubits):
\begin{equation}
	T = tg^p
\end{equation}
where $t$ is the average number of control terminals per qubit and $p$ is the Rent exponent that lies in the range $(0,1]$.
Without any quantum hardware optimisation, the number of terminals will keep increasing linearly with the number of qubits, creating major interconnect problems in the quantum computing stack~\cite{Franke2019RentsComputing}. \\
\begin{figure*}[htb!]
	\centering
	\includegraphics{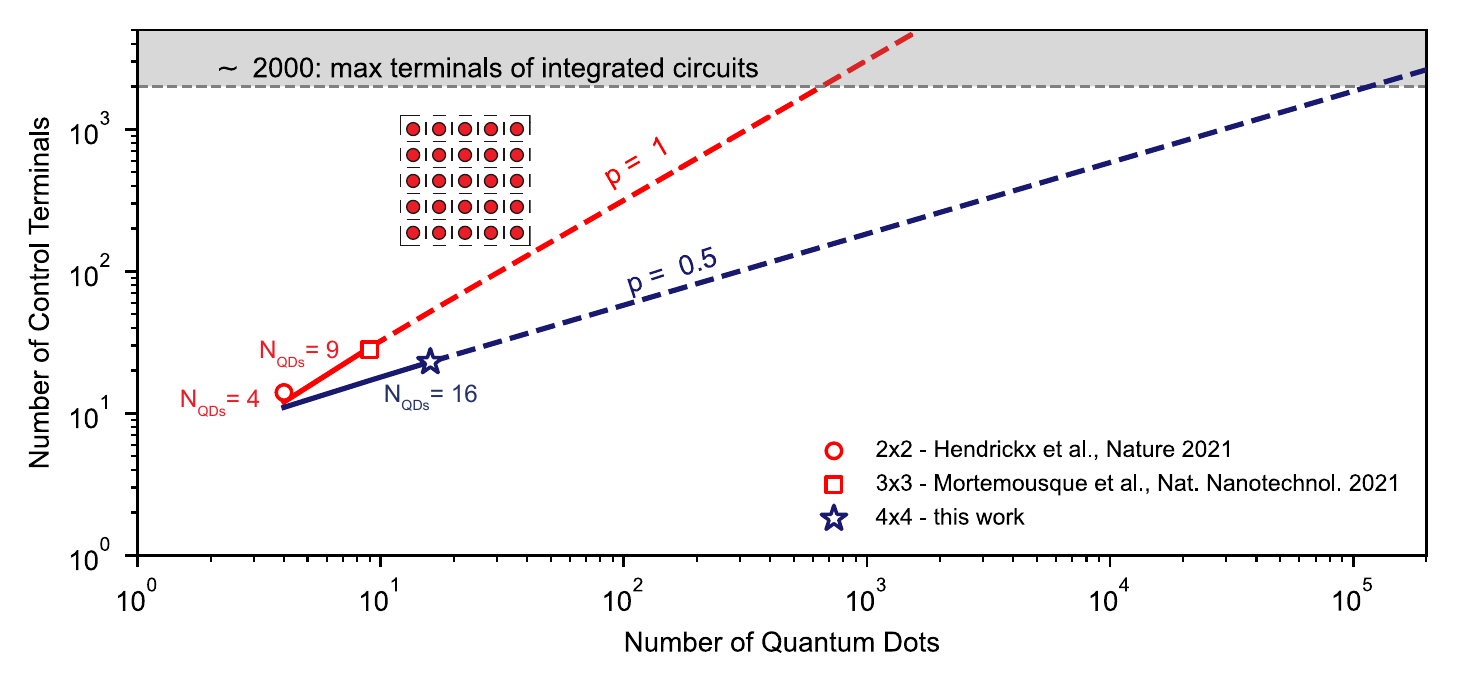}
	\caption{\textbf{Rent's rule in two-dimensional quantum dot architectectures and gates count.} 
		Number of control terminals (gates) versus the number of quantum dots for a two-dimensional array. 
		The blue trace indicates the scaling of our shared-control gate architecture. 
		The red trace shows the scaling for an architecture with individual control (inset) of interdot couplings and quantum dot energies.
		Scatter points represent the gates counts of a 2x2~\cite{Hendrickx2021AProcessor}, 3x3~\cite{Mortemousque2021} and 4x4 quantum dot array (this work).
		We draw a horizontal line at 2000 control terminals, which currently identify the current maximum number of input/output terminals of classical integrated circuits.
		Assuming this as a practical limit for the control terminals of a quantum processor without on-chip control logic~\cite{Vandersypen2017InterfacingCoherent}, the individual control strategy is limited to control up to a few hundred quantum dots.
		In contrast, the shared control approach may be able to control few hundreds of thousands.
	}
	\label{fig:Rent}
\end{figure*}
An analogous challenge emerged in the 1950s in classical electronics when every electrical component needed to be soldered to several others~\cite{Morton1958}.  
The turning point was the invention of the integrated circuit, which led to the realisation of the first microprocessor - the Intel 4004 - with 2300 transistors and only 16 external pins. 
Nowadays, integrated circuits are at the heart of our technology and, with a Rent exponent of about $0.5$ and a maximum number of input/output lines of $T  \sim 2 \cdot 10^3$, present a ratio of transistor to input/output pins $g/T$ of $\sim 10^6$~\cite{Christie2000, Vandersypen2017InterfacingCoherent}. 
On the contrary, in infant quantum chips, the $g/T$ ratio is below 1~\cite{Franke2019RentsComputing}, and therefore, signiﬁcant efforts have to be made to downscale the quantum Rent exponent. \\
In our work, we have moved from electrostatic gating strategies with individual control of every unit to crossbar architectures with shared control of quantum dots energies and interdot couplings~\cite{Li2018}. 
This advance is crucial for scalability because it may enable to use $O(N^{1/2})$ terminals for $O(N)$ qubits (i.e., $p = 0.5$).
The impact of this strategy can be visualised in Suppl. Fig.~\ref{fig:Rent}, where we compare the different scalings of control terminals for a two-dimensional quantum dot array controlled with a shared- and with an individual-control approach.
For the shared-control architecture presented in this work, the gates count as a function of quantum dots $g$ is given by:
\begin{equation}
	T = 6 g^{1/2} - 1 \propto 6 g^{1/2}
\end{equation}
while for an architecture with individual control of all the on-site dot energies and interdot couplings (see inset of Suppl. Fig.~\ref{fig:Rent}), we obtain:
\begin{equation}
	T = 3 g + 2 g^{1/2} - 4 \propto 3 g 
\end{equation}
These equations hold for square arrays with a minimum 2x2 size, i.e., $g \geq 4$. \\
We emphasise that in these considerations we do not account for the terminals controlling the read-out charge sensors.
In the future, a strategy for the integrating charge sensors within quantum dot crossbar arrays needs to be fully work out to establish and operate scalable modules of semiconductor qubit arrays with a Rent exponent of 0.5.
We note that germanium can make ohmic contacts to metals, thus facilitating a very small footprint for charge sensors, and providing a route toward the integration of charge sensors in the quantum dots array.
\section{Typical charge sensors response}
\begin{figure*}[htbp]
	\centering
	\includegraphics{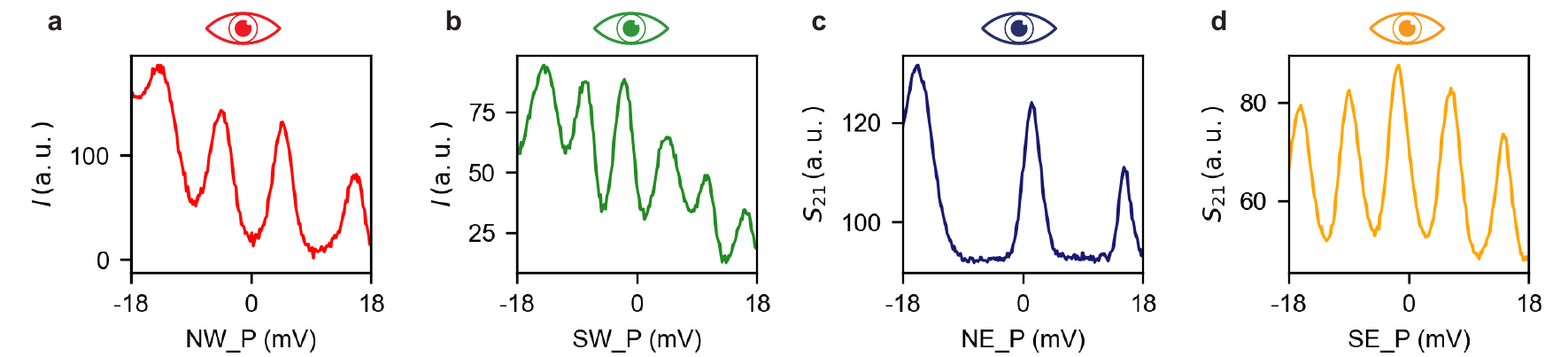}
	\caption{\textbf{Typical charge sensor responses.} \textbf{a, b} Direct-current $I$ via the NW and SW sensor. \textbf{c, d} Reflectometry signal $S_{21}$ of the NE and SE sensor respectively.}
	\label{fig:sensor_traces}
\end{figure*}

To read out the charge states of the 16 quantum dots, we prepare four charge sensors in the Coulomb regime, as demonstrated by the transport features in Suppl. Fig.~\ref{fig:sensor_traces}, and operate them at the steepest points. 
\newpage
\section{Virtual gate matrix}
\begin{figure*}[htbp]
	\centering
	\includegraphics{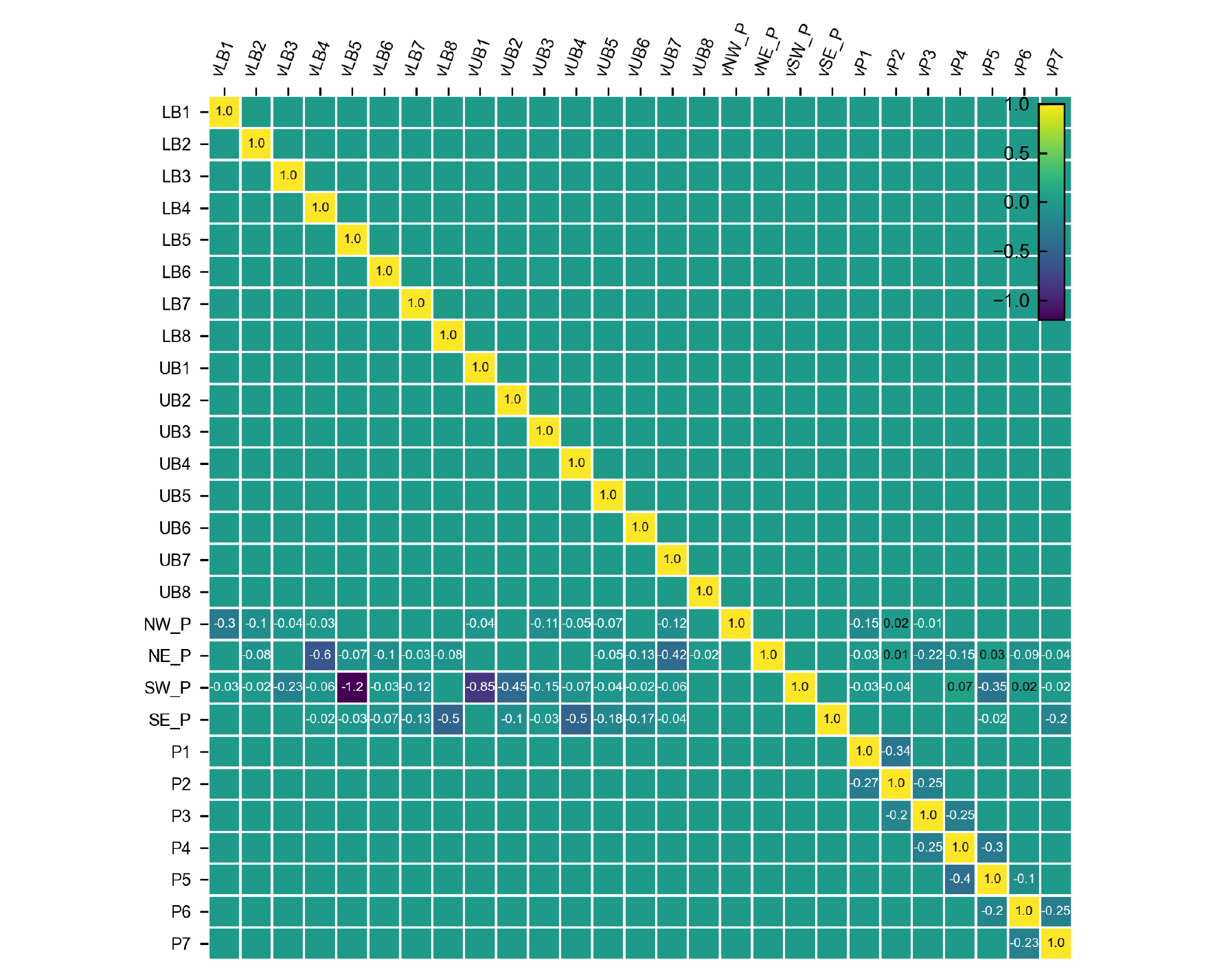}
	\caption{\textbf{Virtual gate matrix.} 
		Visualization of the virtual gate matrix. Virtual barriers vLB$i$ and vUB$i$ are defined as combination of the relative barriers LB$i$ and UB$i$ with $i \in [1, 8]$ and sensor plungers.
		NW\textunderscore P, NE\textunderscore P, SW\textunderscore P and SE\textunderscore P identify the plungers of the respective charge sensors.}
	\label{fig:virtual_matrix}
\end{figure*}

In Suppl. Fig.~\ref{fig:virtual_matrix} we display the virtual matrix used in software to define the set of virtual gates.
We define the virtual barriers such that their variation does not influence the position of the charge sensors Coulomb peak.
Virtual plungers are designed to be able to tune independently the on-site energies of the quantum dots by using only nearest-neighbours compensations.
Because multiple sites are controlled by a single plunger line and each of the site can have a slightly different crosstalk to the surrounding gate, our procedure results in an approximation. 

\clearpage
\section{Practical device improvements} 
\label{sec:improvements}
Here, we present a series of practical ideas to mitigate challenges observed (and not) throughout the experiment.
\begin{enumerate}
	\item The tune-up of the current implementation is complicated by the emergence of a few ($\sim 5$) unwanted stray dots outside the array, located under the dense gates fanout.
	These decrease the read-out sensitivity to the designed quantum dots and, in general, complicate the system tune-up. \\
	Near-term solutions can be: the addition of a lower ``screening'' gate layer (kept at a more positive voltage), or an upper ``depleting'' gate layer that prevents any charge puddles from forming in between different gate lines. 
	A more sophisticated solution consists of the implementation of vertical interconnect access through the oxide layer that enables to fan out the gate lines at a much higher level in the stack~\cite{Ha2022}.
	We also envision that, in the future, boundary effects of large qubit arrays need to be accounted for (e.g., dots at the perimeter of the array experience a smaller electric field than the ones in the middle). 
	Hence, another possibility is to neglect the dots at the perimeter completely.
	
	\item A faster device tunability can be achieved by using radio frequency reflectometry, which, in our experiment, worked out only partially.
	
	\item A higher level of homogeneity and functionality than is presented here will be needed for practical quantum computation with quantum dots.
	Leaving aside the development on the material stack itself (Ge/SiGe) and on the device nanofabrication, there appears to still be space for improvements on the gate layout.\\
	Similarly to the behaviour of the turn-on voltages in transistors~\cite{Bavdaz2022}, a higher quantum dots homogeneity may be achieved by increasing their size.  
	This will also have the beneficial effect of increasing the plunger gates lever arm, which will be less screened by the lower layers.
	Clearly, an excessively large quantum dot can lead to (Pauli spin-blockade) read-out problems if the energy of the first excited state is too low. 
	Therefore a search for an optimal size needs to be performed.
	
	\item In the current device, we observed that the ratio of the two barrier layers lever arms is approximately 2-3. 
	Due to imperfections in the nanofabrication and an oxide layer in between them, the top layer partially overlaps the bottom one. 
	As a consequence, its electric field is screened. To enhance the coupling of the top layer, one can design the top layer to have a slightly larger width, or add a gap of a few nm in between the two lines.	 
\end{enumerate}

\clearpage
\section{The  quantum dots identification}
\begin{figure*}[htbp]
	\centering
	\includegraphics{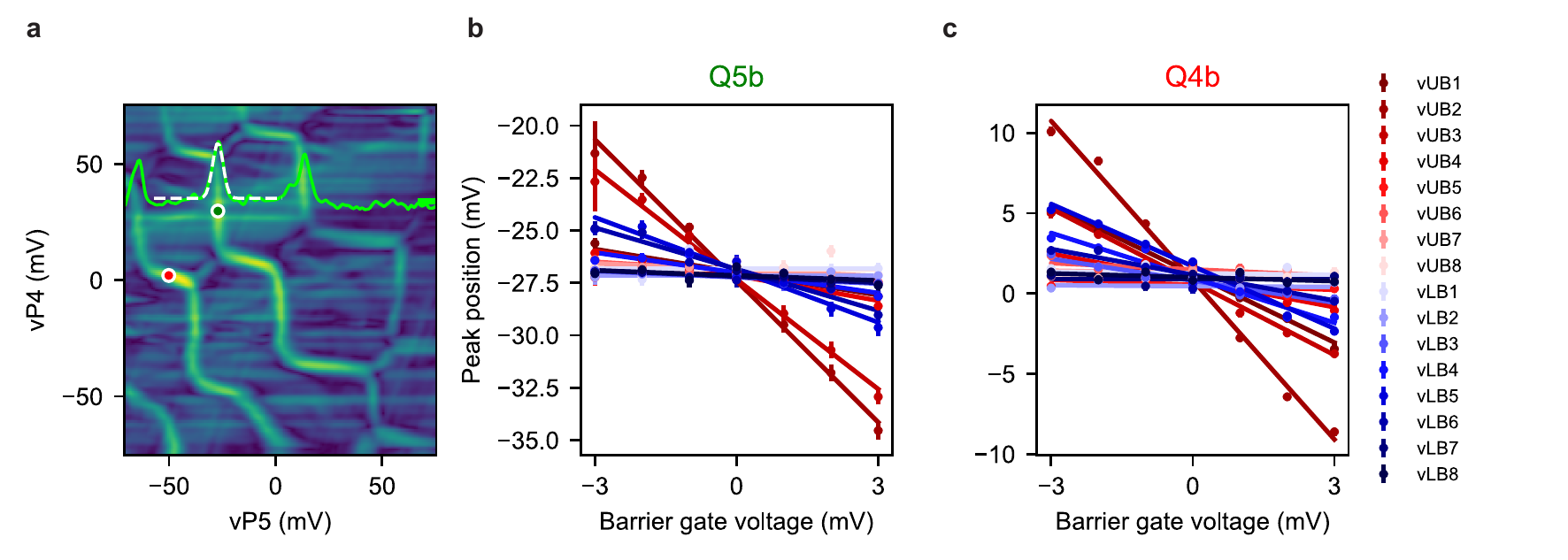}
	\caption{\textbf{Detecting the shift of the quantum dots transition lines: an exemplary case.}
		\textbf{a,}~Charge stability diagram (gradient of the SW sensor signal) showcasing a set of vertical and horizontal charge transition lines. 
		The green trace shows the derivative of the data at the vP4 value identified by the horizontal green tick.
		The white dashed line is a fit of the linecut with a Gaussian function, and the green marker identifies the fitted centre of the peak. 
		The red marker labels the fitted coordinate of a vertical transition line (not shown).
		\textbf{b, c,}~Scatter points are the fitted positions of the green and red markers, respectively, as a function of the voltage applied at all barriers. 
		Error bars on the points display the standard deviation of the fitted centre of the Gaussian.
		Dashed lines are the best linear fit to the data. The normalised absolute values of the slope parameter are taken as capacitive couplings of each gate to the specific transition line.
		The error bar on the capacitive coupling is taken as one standard deviation of the fitted slope parameter.
		The horizontal (vertical) transition line in \textbf{(a)} is attributed to the quantum dot Q4b (Q5b). 
	}
	\label{fig:shift_lines}
\end{figure*}

\begin{figure*}[htbp]
	\centering
	\includegraphics{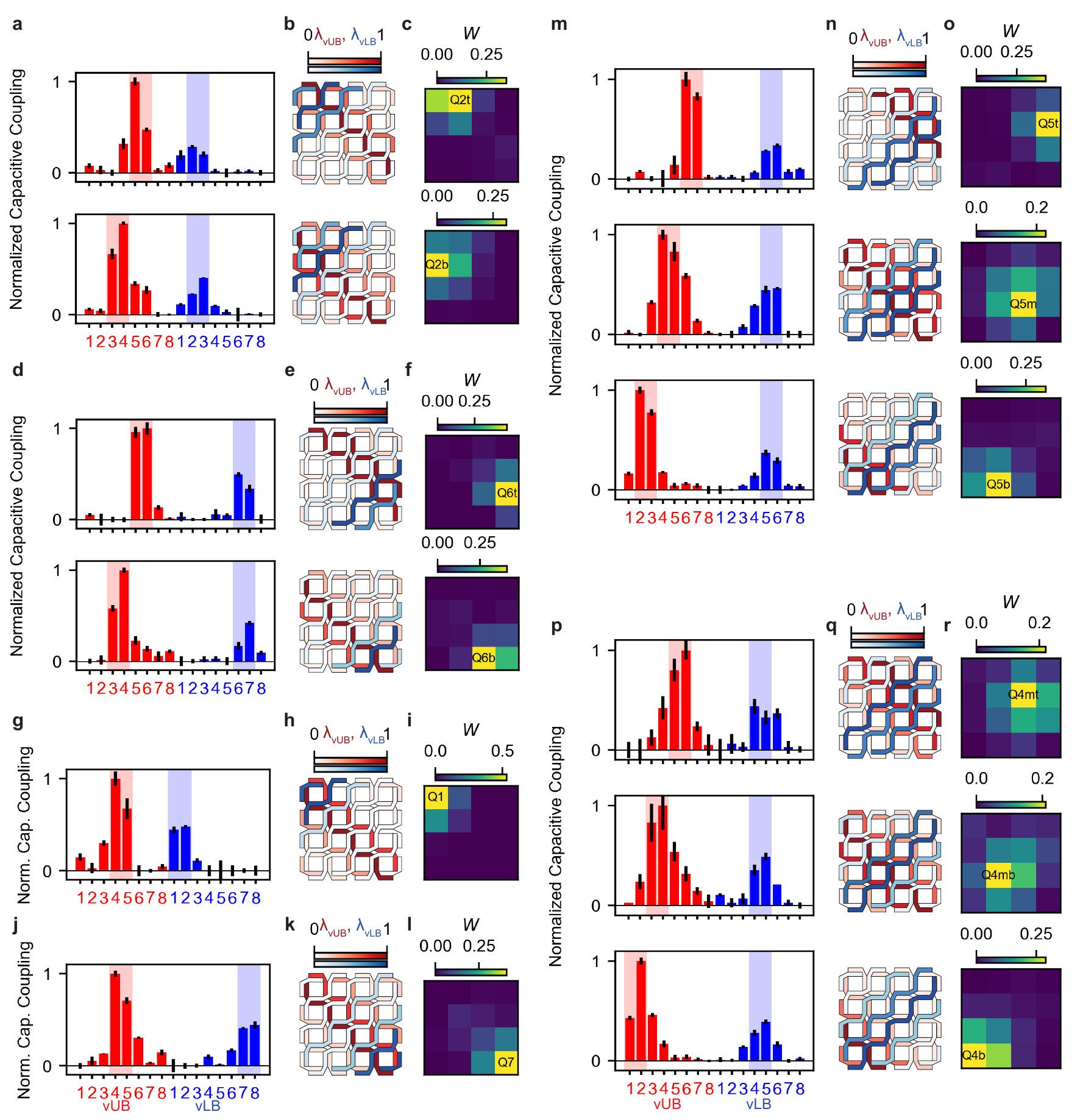}
	\caption{\textbf{Quantum dot identification.}
		\textbf{a,} Histograms of the (normalised) capacitive couplings (in red for the vUB, and in blue for the vLB gates) obtained by the analysis of transition lines attributed to different quantum dots as shown in Suppl. Fig.~\ref{fig:shift_lines}.	
		Red and blue backgrounds are added to emphasise the two barriers that surround the labelled quantum dot. 
		\textbf{b,} Device layout with the capacitive couplings colour-coded on the filling of the gate lines. The vUB (vLB) capacitive couplings are normalised to their maximum values.
		Intuitively, the quantum dot associated with the analysed transition lines is located at the intersection of the two intensely coloured red and blue lines.
		\textbf{c,} Extracted probabilities ($W$) of each set of addition lines (calculated as discussed in the Methods). 
		The comparison of the top and bottom panels of \textbf{a, b, c} (\textbf{d, e, f})  clearly distinguishes the two Q2 (Q6) quantum dots.
		Similarly, our method is applied on the two dots controlled by independent plunger lines Q1 and Q7 in panels \textbf{g, h, i} and \textbf{j, k, l}, respectively.
		The three different rows of panels \textbf{m, n, o} (\textbf{p, q, r}) enables to label the Q5 (Q4) quantum dots.
		The dot Q4t could not be systematically analysed because, in this gating regime, we observe a slow loading mechanism via the defective barrier UB8 with respect to the timescale of our scan ($\sim $ ms). 
		However, because such transition lines are controlled by vP4 (labelled in Suppl. Fig.~\ref{fig:first_16_QDs}), do strongly respond to vUB7 and vLB5, we can still map them to the site Q4t, in a qualitative way. 	
	}
	\label{fig:QDs_identification}
\end{figure*}

To obtain the capacitive coupling of each barrier to each dot, we analyse several charge stability diagrams as shown in Suppl. Fig.~\ref{fig:shift_lines}a.
We monitor the position of the transition lines (away from charge interdots) by fitting the derivative of the data with a Gaussian function, after subtraction of a slow-varying background.
In the small voltage range that we consider, the extracted peak positions respond linearly to each barrier gate (Suppl. Figs.~\ref{fig:shift_lines}b, c)
The normalised slope of the fitted linear function is used to quantify the capacitive coupling of each gate.
The latter are plotted as histograms and visualised on the device layout in Suppl. Fig.~\ref{fig:QDs_identification}.
As described in the Methods, this information can be used to clearly assign each set of transition lines to the dot in the grid.
\clearpage
\section{Tune-up of the crossbar array in the few-hole regime}
\begin{figure*}[htbp]
	\centering
	\includegraphics{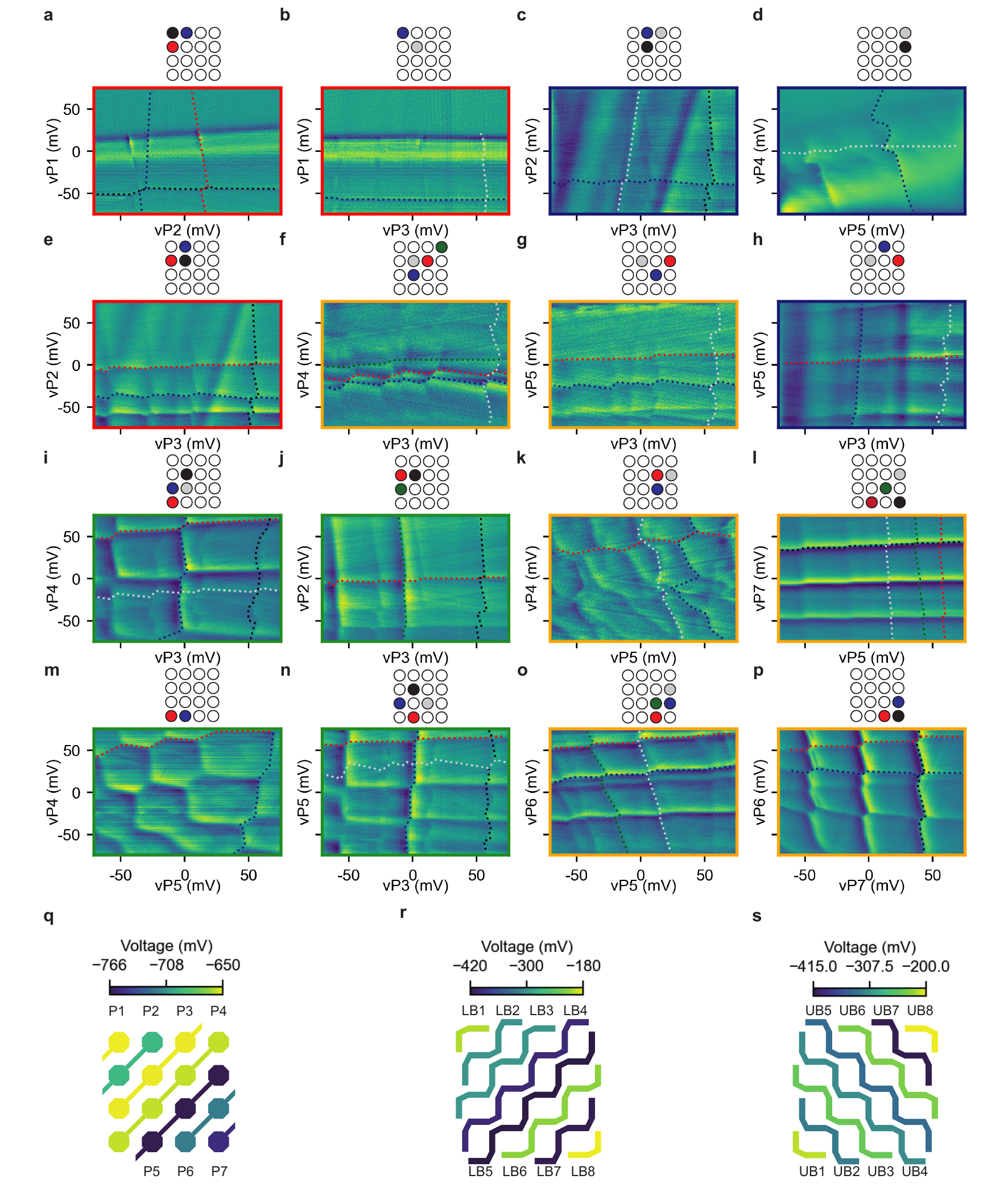}
	\caption{\textbf{Tune-up of the crossbar array in the few-holes regime.}
		\textbf{a-p,}~Charge stability diagrams (raw sensor signal after subtraction of a background) showcasing the 16 quantum dots system in the few-hole regime.
		These measurements are taken in the first tune-up of the crossbar array.
		In each map, the first visible transition lines from the right or the top are labelled and assigned to the relative quantum dot by dashed lines with colours defined in the quantum dots schematic at the top. 
		The quantum dots identification is performed via the results of Suppl. Fig.~\ref{fig:QDs_identification}.
		The colour of the panel frame identifies the sensor used: NW in red, NE in blue, SW in green, and SE in ochre. 
		\textbf{q-s,}~Schematics of the voltage applied at all crossbar gates at this phase of the experiment.    
		The voltages are here optimised to circumvent stray dots.
	}
	\label{fig:first_16_QDs}
\end{figure*}

In Suppl. Fig.~\ref{fig:first_16_QDs}, we display the charge stability diagrams obtained for the first tune-up of the device with the quantum dots in the few-holes regime. 
These maps are part of the data sets that are used for the analysis presented in Suppl. Fig.~\ref{fig:QDs_identification}.
\newpage
\clearpage
\section{Addressable exchange operations with a double barrier design}
\begin{figure*}[htb]
	\centering
	\includegraphics{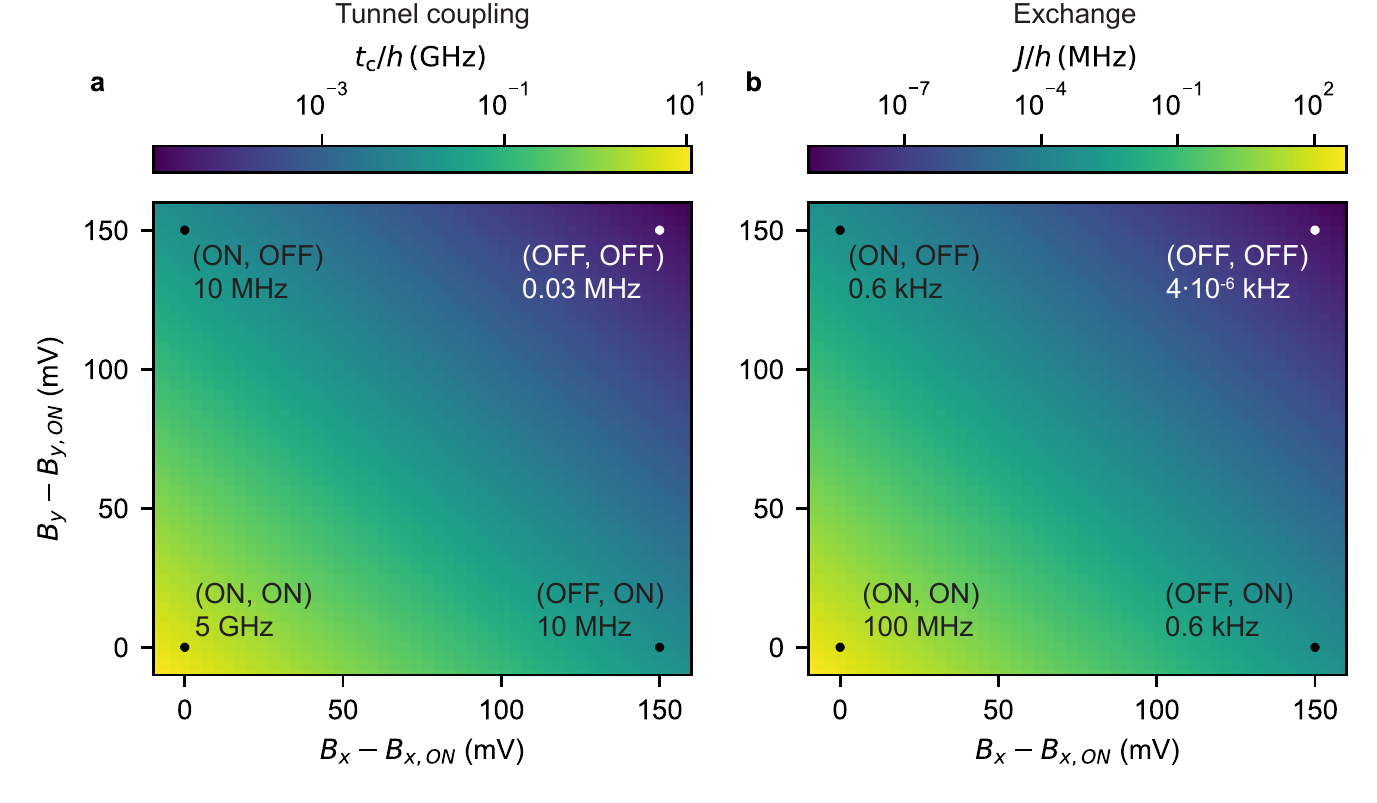}
	\caption{\textbf{Addressable exchange operation with a double barrier design.}
		\textbf{a, b} Target tunnel coupling and exchange required for fast two-qubit gates with interaction ON (bottom left corner), and OFF (top left, top right, bottom left corners in the map).
		The exchange interaction of qubit pairs at (ON, OFF) crossing points remains four orders of magnitude smaller than in the (ON, ON) cases.
		The parameters displayed are calculated at the coordinates (0, 0), (0, 150), (150, 0), (150, 150) mV.
	}
	\label{fig:addressable_exchange}
\end{figure*}

The ability to control the tunnel coupling with two barriers opens the opportunity to design addressable exchange-based two-qubit gates in architectures with shared control.
We envision an operation strategy in which a two-qubit gate can be activated only when both barriers are in the ON state. 
For a fast CPHASE gate with a duration of 5 ns, we require to activate an exchange interaction of $J_\mathrm{ON}/h = 100$ MHz. 
In all other cases (i.e., when the barriers are in the configurations (ON, OFF), (OFF, ON), (OFF, OFF)), we demand a sufficiently low exchange to minimise errors. 
State-of-the-art values of OFF exchange interaction are in the order of $J_\mathrm{OFF}/h \sim $ 10 KHz~\cite{Xue2022}. \\~\\
Suppl. Fig.~\ref{fig:addressable_exchange} illustrates the required barrier voltage points to obtain such exchange interactions considering symmetric lever arms, a quantum dot charging energy of U = 1 meV and operations at the charge symmetry point (i.e., at zero detuning).
In Suppl. Fig.~\ref{fig:addressable_exchange}, we have approximated the dependence of the tunnel coupling energy $t_\mathrm{C}$ with respect to the two barrier voltages $B_{x} $ and $B_{y}$ with ~\cite{Hendrickx2019_2Qubits, Russ2018, Loss1998QuantumDots}:
\begin{equation}
	t_\mathrm{C} =   \frac{\sqrt{J U}}{2} = c_1 \cdot e^{-c_2 \alpha (B_{x} - B_{x, ON})} \cdot e^{-c_2 \alpha (B_{y} - B_{y, ON})}
\end{equation}
with $B_{x, ON}$ and $B_{y, ON}$ the ON set-points of the two barriers. \\
In this example, we have set the prefactor $c_1$ to 5, and the effective barrier lever arms $c_2 \cdot \alpha$ to 0.04 $V^{-1}$.
\clearpage
\section{Two-axis control of the quantum dots interdot transition line}
\begin{figure*}[htbp]
	\centering
	\includegraphics{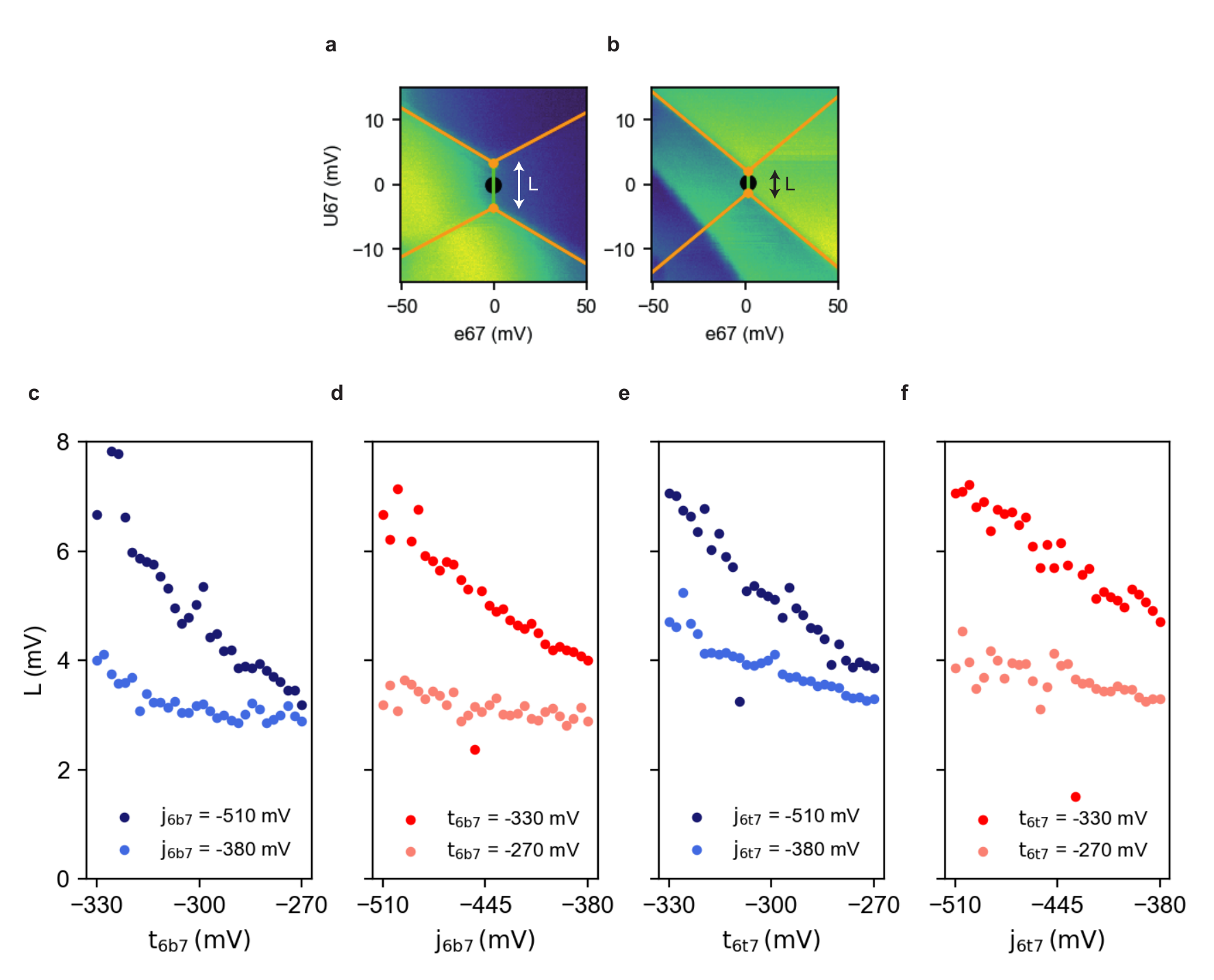}
	\caption{\textbf{Two-axis control of the quantum dot interdot transition line.}
		\textbf{a, b,} Exemplary charge stability diagrams taken at the Q6t-Q7 (3,1)-(2,2) charge interdot line. 
		The two maps are taken at the diagonally opposite points of the two-dimensional barrier scan, at $t_\mathrm{6t7}, j_\mathrm{6t7}$ = (-330, -510) and (-270, -380) mV, respectively.
		Orange lines represent a fit to the map following the procedure shown in ref.~\cite{VanDiepen2018}. A black circle identifies the fitted centre of the interdot, and the arrow illustrates the size in voltage of the interdot line.
		\textbf{c, d,} Size of the Q6b-Q7 (3,1)-(2,2) charge interdot line as a function of the two virtual barriers.
		\textbf{e, f,} Same for the Q6t-Q7 (3,1)-(2,2) charge interdot line.
		Outliers in the plot are due to non-accurate fits of the image.
	}
	\label{fig:interdot}
\end{figure*}

By fitting the two-dimensional (e67, U67) charge stability diagrams (examples in Suppl. Figs.~\ref{fig:interdot}a, b), we obtain an estimate of the (3,1)-(2,2) charge interdot size $L$ for the double dot systems Q6b-Q7 and Q6t-Q7. 
The size of the interdot line is indicative of the capacitive coupling between the adjacent quantum dots. 
Consistent with the two-axis tunability of the tunnel coupling, the interdot size is varied as the effective distance between the dots is modified by the action of the two tunnel barriers (Suppl. Figs.~\ref{fig:interdot}c-f).
\section{Two-axis control of the Q6b-Q5m interaction}
\begin{figure*}[htbp]
	\centering
	\includegraphics{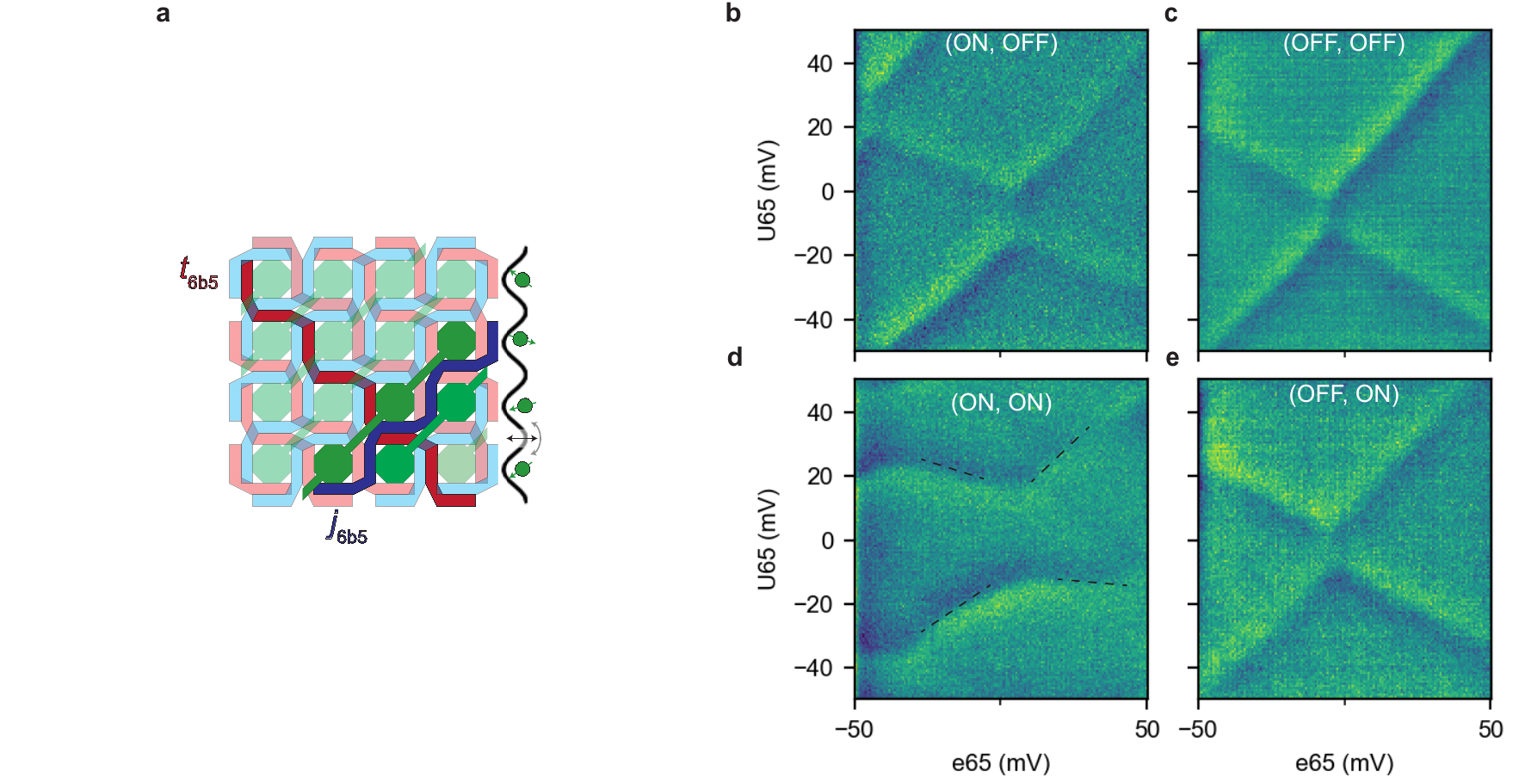}
	\caption{\textbf{Two-axis control of the Q6b-Q5m interdot coupling.} \textbf{a} Schematic of the crossbar indicating the two intersecting virtual barriers (in red and blue) controlling the Q6b-Q5m interaction.
		\textbf{b-e} Charge stability diagrams at different barrier voltages: \textbf{(b)} $(t_{\mathrm{6b5}}, j_{\mathrm{6b5}}) =$ (-330, -380) mV, \textbf{(c)} (-270, -380) mV, \textbf{(d)}  (-270, -510)  mV, \textbf{(e)}  (-330, -510)  mV. 
		In \textbf{(d)}, in the high interdot coupling regime, we add dashed lines as guide for the eyes on weakly visible transition lines. 
		Here, we display the signal from the SE charge sensor after subtraction of a background.
	}
	\label{fig:P6b-P5m_interdot}
\end{figure*}

We repeat the tunnel coupling experiments considering the double-dot pair Q6b and Q5m by defining the virtual barriers $t_\mathrm{6b5}$ and $j_\mathrm{6b5}$ starting from the relative UB4 and LB6 barriers (Suppl. Fig.~\ref{fig:P6b-P5m_interdot}):
{\begin{equation*}
		\begin{pmatrix}
			\text{P5} \\ \text{P6} \\ \text{UB4}\\ \text{LB6} \\ \text{SE\textunderscore P}
		\end{pmatrix}
		= 
		\begin{pmatrix} 
			-1.63 & -0.58 \\
			-1.61 & -0.48 \\
			1 & 0\\
			0 & 1 \\
			-0.43 & -0.02 \\
		\end{pmatrix}
		\begin{pmatrix} 
			\mathrm{t_{6b5}} \\
			\mathrm{j_{6b5}} \\
		\end{pmatrix}
\end{equation*}}
We observe that the limited sensitivity at the interdot hinders a quantitative tunnel coupling analysis. However, at a qualitative level, we still observe the expected pattern with the size of the interdot tunable by both barriers. \\
\clearpage
\section{Electron temperature extraction}
\begin{figure*}[htp]
	\centering
	\includegraphics{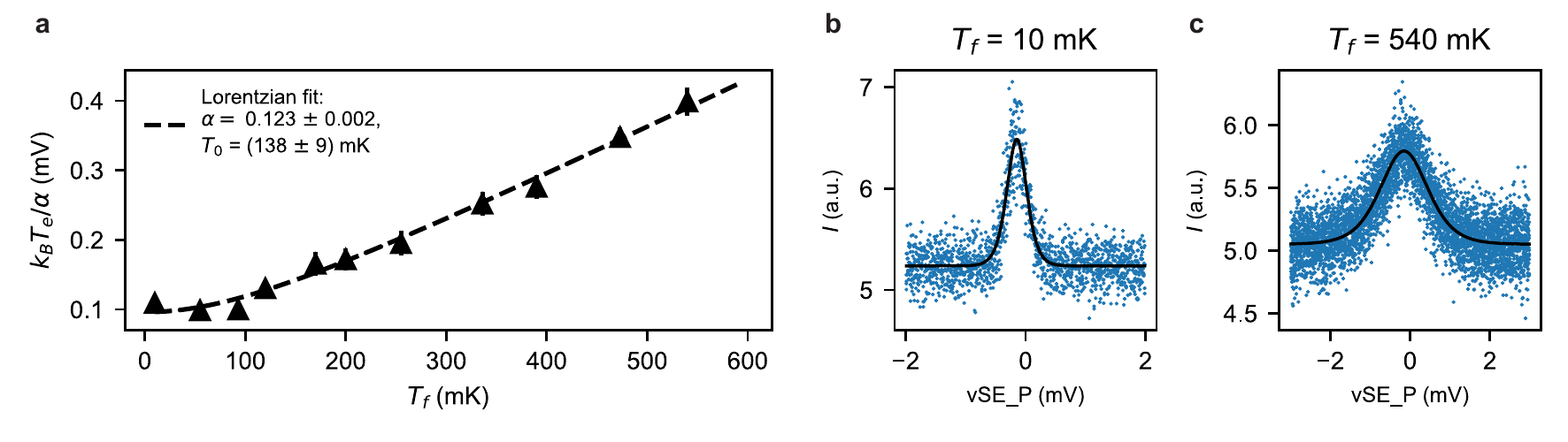}
	\caption{\textbf{Electron temperature extraction.}
		\textbf{a} Coulomb peak width as a function of fridge temperature.  
		\textbf{b, c.} Exemplary data (scatter points) and best fits (black lines) collected at $T_f=10$ mK and $T_f=540$ mK, respectively. 
		For every temperature, we perform five plunger sweeps across the Coulomb peak (each averaged 500 times) and, by fitting the curves, we obtain five different estimates for the Coulomb peak width. 
		In \textbf{(a)}, we show the average width. The depicted standard deviation is determined by considering the standard deviation from the single fit itself or the standard deviation between the five different fitting estimates, depending on which is dominant.
	}
	\label{fig:electron_temperature}
\end{figure*}
We estimate the electron temperature of our setup by fitting a temperature-broadened Coulomb peak of the SE charge sensor.
We choose a source-drain voltage $V_{\mathrm{SD}}$ and coupling energy to the leads $h \Gamma$ such that $h\Gamma, e V_{\mathrm{SD}} << k_B T_e$, with $h$ and $k_B$ the Plank and Boltzmann constants, respectively and $\Gamma$ the lead-dot tunnel rate.
The current $I$ at the Coulomb peak is fitted with the Lorentzian distribution
\begin{equation}
	I_{model} = a \cdot \cosh^{-2} \left( \frac{\alpha_{sensor} \cdot \epsilon}{2k_B T_e} \right) + c,
\end{equation}
where $a$ is an amplitude prefactor, $\alpha_{sensor}$ is the lever arm of the sensor plunger gate, $\epsilon$ is the plunger gate voltage and $c$ an offset. 
The full width half maximum (FWHM) of the peak is therefore given by
\begin{equation}\label{FWHM}
	\text{FWHM} = \frac{k_B T_e}{\alpha_{sensor}} = \frac{k_B \sqrt{T_0^2 + T_f^2}}{\alpha_{sensor}},
\end{equation}
where $T_f$ is the nominal fridge temperature and $T_0$ the base electron temperature~\cite{Petit2018SpinQubits}. 
We use the FWHM dependence on the nominal fridge temperature to extract both $T_0$ and $\alpha_{sensor}$.
The result can be seen in Suppl. Fig.~\ref{fig:electron_temperature} with two exemplary fits for 10 and 540 mK.
The fit with \autoref{FWHM} results in $T_0 = 138 \pm 9$ mK and $\alpha_{sensor} = 0.123 \pm 0.002$ eV/V, which is consistent with an independent lever arm extraction from Coulomb diamonds. 

\section{Detuning lever arm extraction}
\begin{figure*}[htbp]
	\centering
	\includegraphics{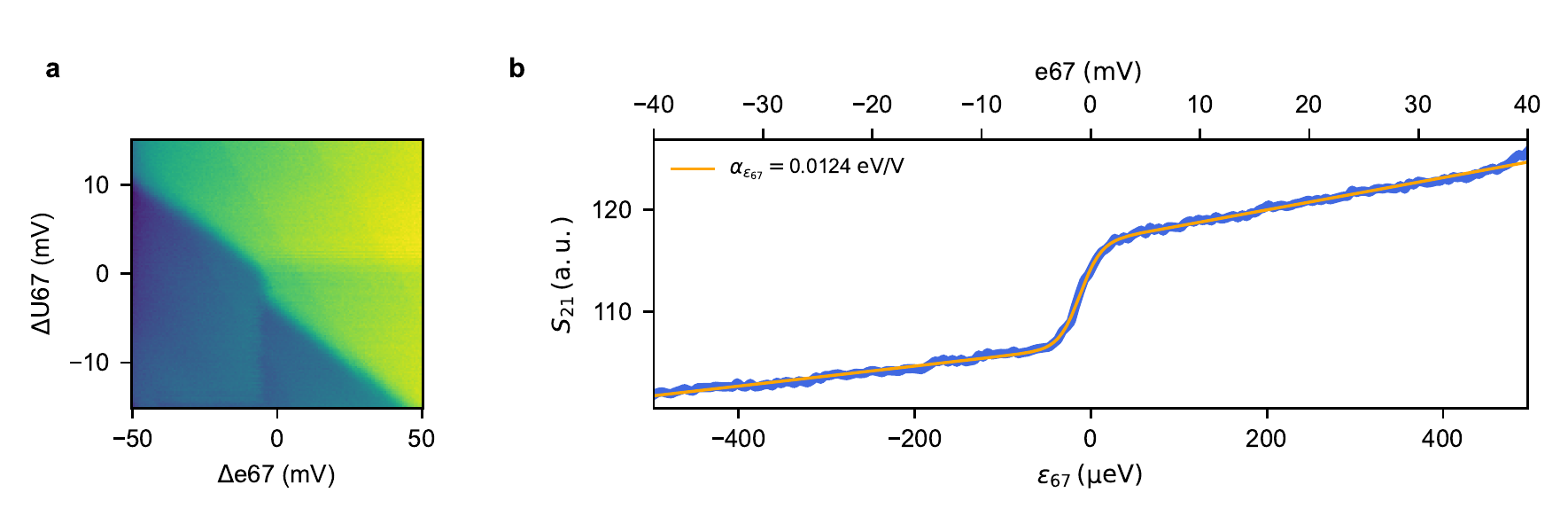}
	\caption{\textbf{Temperature limited polarisation line.}
		\textbf{a,} Charge stability diagram of the uncoupled Q6b-Q7 at the (3,1)-(2,2) charge interdot. 
		In this configuration, the tunnel time in Q6b is small with respect to the ramp time of the detuning axis, therefore the interdot is extended and the Q6b addition line is not clearly visible in the map.
		\textbf{b,} Thermally limited polarisation line (blue trace) taken at the centre of the interdot shown in \textbf{(a)}. From the best fit (orange line), we obtain the detuning lever arm $\alpha_{\epsilon_{\mathrm{67}}} = 0.012(4)$ eV/V.}
	\label{fig:detuning_lever_arm}
\end{figure*}

For an accurate estimation of the interdot tunnel coupling, we evaluate the quantum dot detuning lever arm by fitting the sensor signal $S_{21}$ with a thermally limited polarisation line (Suppl. Fig.~\ref{fig:detuning_lever_arm}) as described in \cite{DiCarlo2004DifferentialDot}:
\begin{equation} \label{equ:sensor_conductance}
	S_{model} = S_0 \pm \delta S \frac{\epsilon}{\Omega} \cdot \tanh \left( \frac{\Omega}{2 k_B T_e} \right) + \frac{\partial S}{\partial \epsilon} \epsilon
\end{equation}
with $S_0$ the background signal of the charge sensor, $\delta S$ the signal amplitude, $\epsilon$ the detuning energy,  $\Omega$ the energy difference between the two levels and the term $ \frac{\partial S}{\partial \epsilon} \epsilon $ a linear slope due to cross-talk to the charge sensor.
In the low-tunnelling regime, we can approximate $\Omega = \sqrt{\epsilon^2 + 4t^2} \approx \epsilon$, which reduces \autoref{equ:sensor_conductance} to
\begin{equation}
	S_{\mathrm{model}} = S_0 \pm \delta S \cdot \tanh \left( \frac{\alpha_{QD} \cdot \epsilon}{2 k_B T_e} \right) + \frac{\partial S}{\partial \epsilon} \epsilon.
\end{equation}

\clearpage
\section{Gate voltages of the crossbar in the odd charge regime}
\begin{figure}[htbp]
	\centering
	\includegraphics{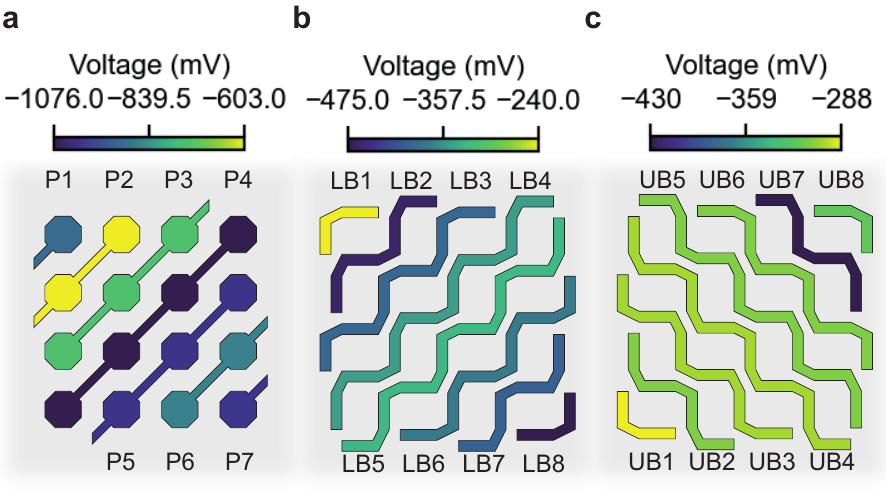}
	\caption{\textbf{Crossbar gate voltages.} \textbf{a-c,} Voltages applied at the real P, LB and UB gates, respectively, when the system is tuned in the odd charge occupation regime. 
	}
	\label{fig:voltages_gates}
\end{figure}

\section{Variability of addition voltages}
\begin{figure*}[htbp]
	\centering
	\includegraphics{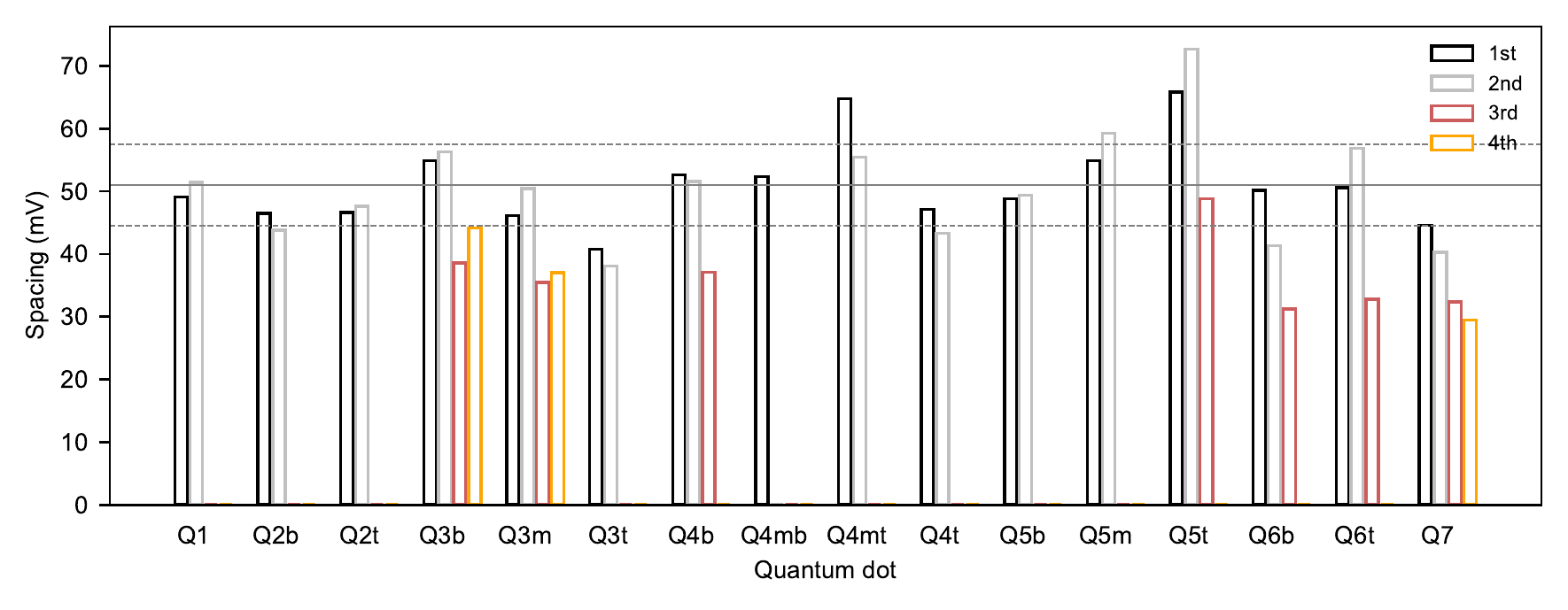}
	\caption{\textbf{Charge dddition voltages of all the quantum dots.} Histogram of the addition voltages of each quantum dot considering the first spacings, when detectable in the already existing datasets.
	The three horizontal lines identify the spread of the first hole addition voltage of $51 \pm 6$ mV.
	}
	\label{fig:gate_spacing}
\end{figure*}
We extract the charge addition voltages of all the quantum dots from the corresponding stability diagrams to evaluate the level of homogeneity of the system (Suppl. Fig.~\ref{fig:gate_spacing}).
We consider the charge addition voltage as the spacing between two consecutive charge addition lines.
We estimate the averaged spacing of the first and second hole to be $51 \pm 6$ mV and $50 \pm 9$, respectively, indicating a $\sim 10-20\%$ variability of the quantum dot properties in the array. 
\clearpage
\section{Demonstration of the odd occupation regime }
\begin{figure*}[htb]
	\centering
	\includegraphics{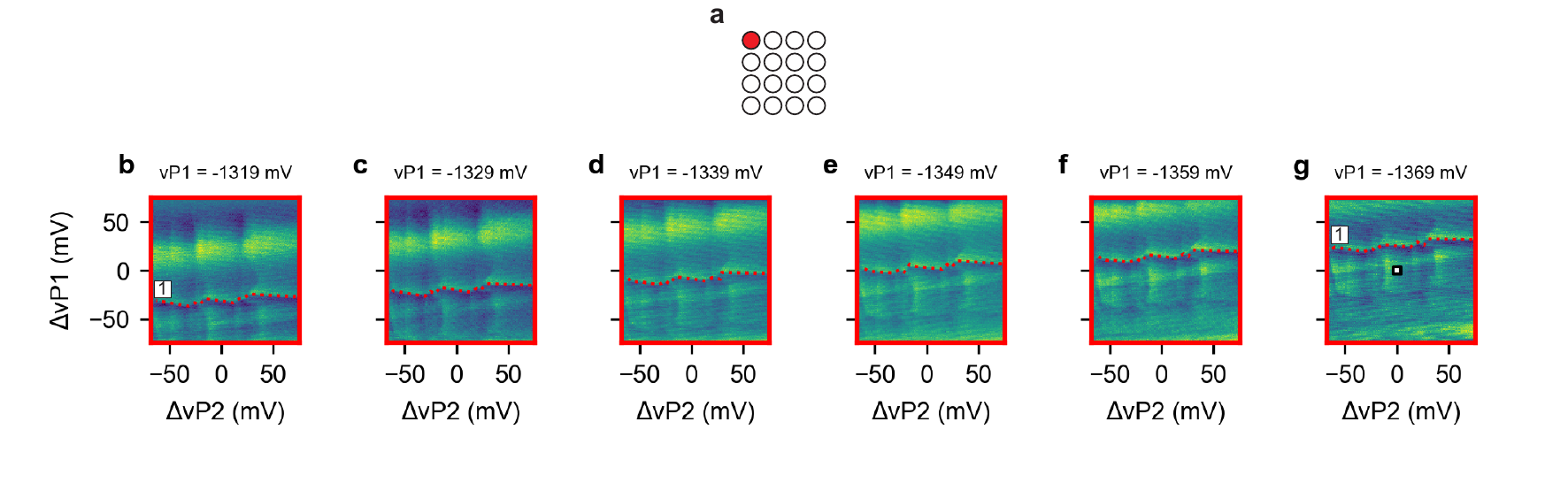}
	\caption{\textbf{Counting the number of holes in Q1.}
		\textbf{a,} Schematic of the quantum dot grids labelling the dot Q1	
		\textbf{b-g,} Sequence of measurements (for improved visibility, we show here the raw data after subtracting a slow-varying background) that demonstrates the presence of a single hole in Q1.
		We use a dashed guide-for-the-eye trace and a label to indicate the first Q1 addition line.}
	\label{fig:odd_P2P1_1}
\end{figure*}

\begin{figure*}[htbp]
	\centering
	\includegraphics{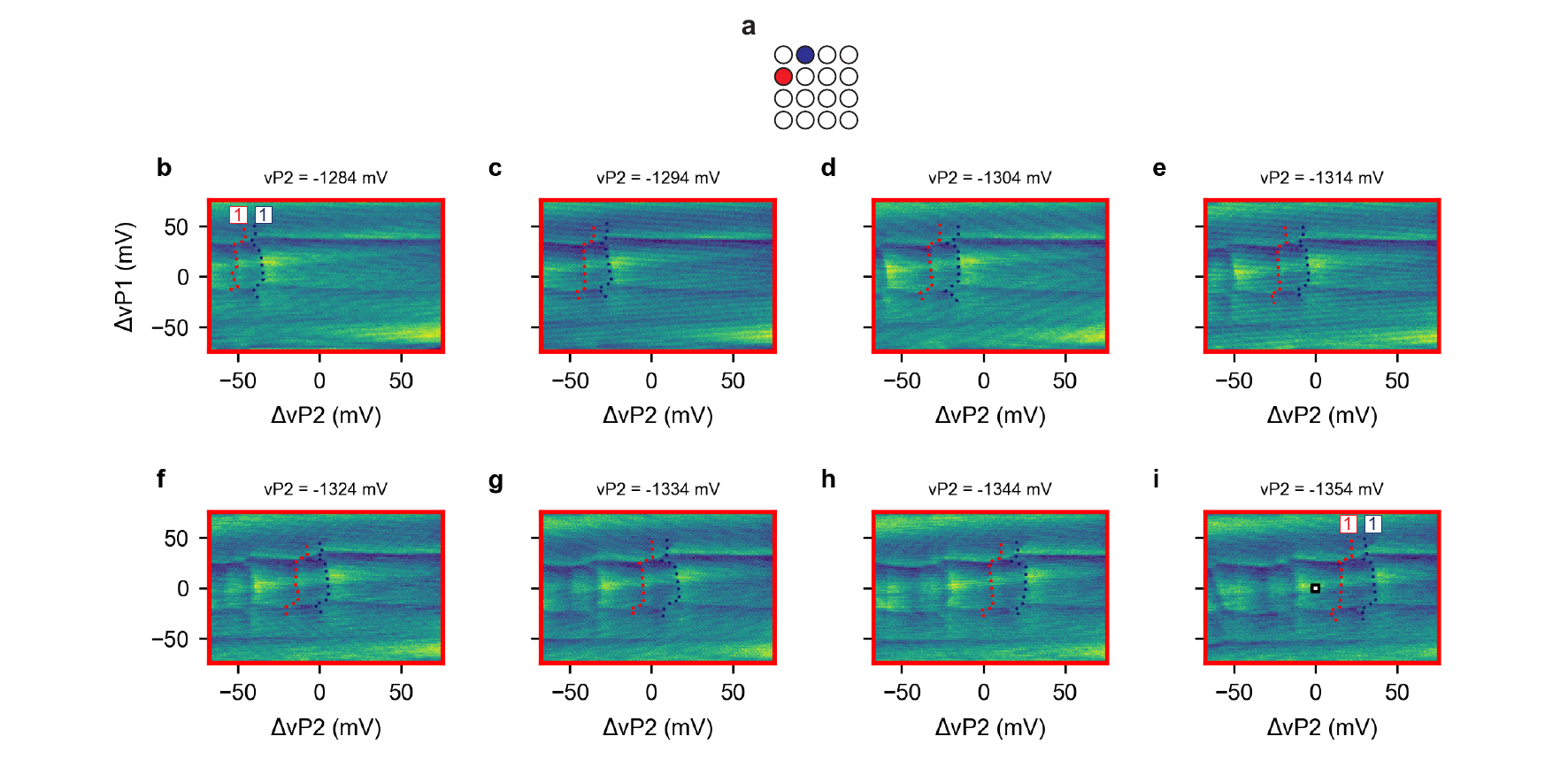}
	\caption{\textbf{Counting the number of holes in Q2b and Q2t.} 
		\textbf{a,} Schematic of the quantum dot grids labelling the dots Q2b and Q2t in different colours.	
		\textbf{b-e,} Sequence of measurements (for improved visibility, we show here the raw data after subtracting a slow-varying background) that demonstrates the single-hole occupancy of Q2b and Q2t.
		In \textbf{(e)}, the square identifies the (Q1, Q2b, Q2t) (1,1,1) charge state. The coloured dashed lines and the label identify the two first addition lines.
	}
	\label{fig:odd_P2P1_1_P2s}
\end{figure*}

\begin{figure*}[htbp]
	\centering
	\includegraphics{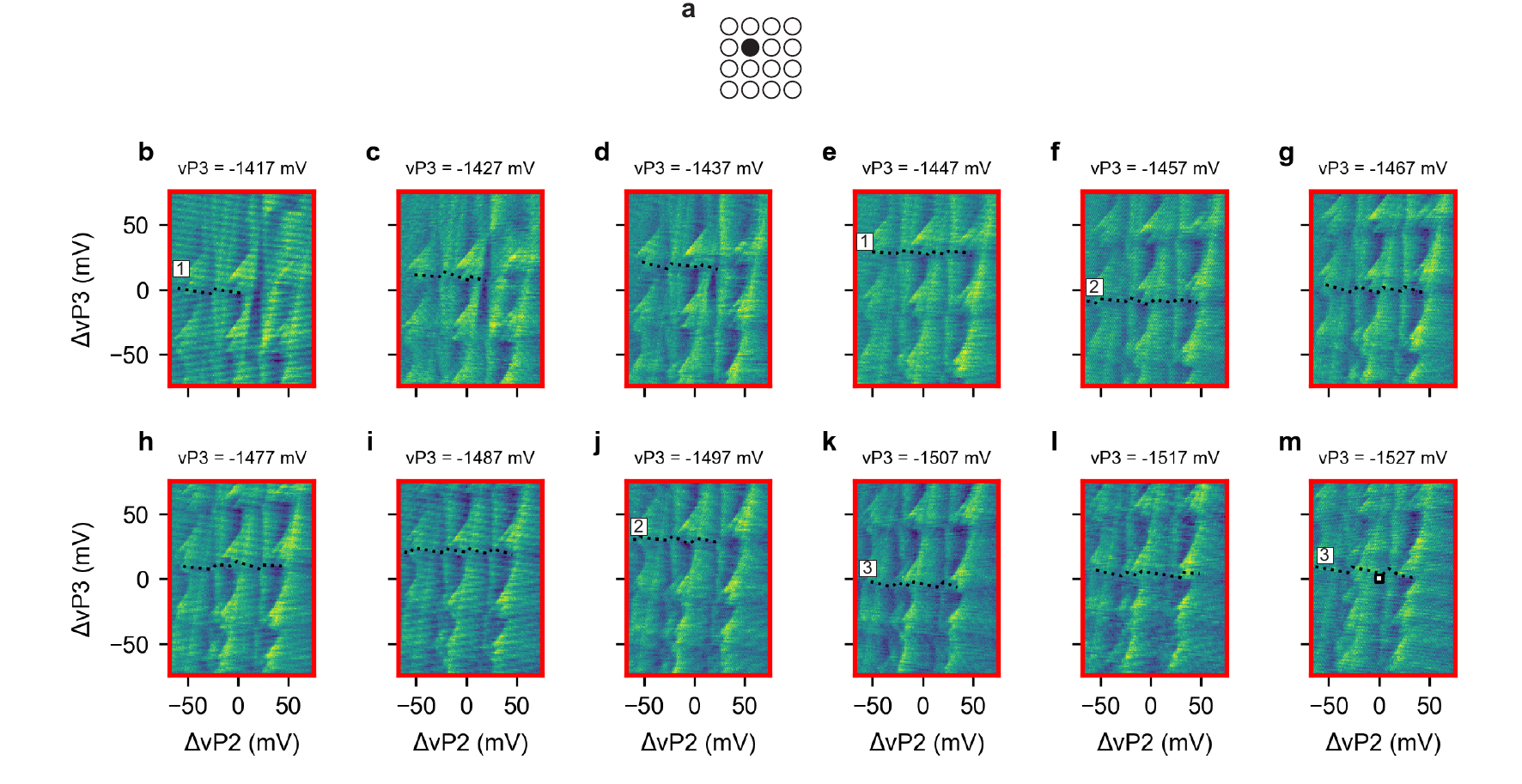}
	\caption{\textbf{Counting the number of holes in Q3m.} 
		\textbf{a,} Schematic of the quantum dot grids labelling the dot Q3m.	
		\textbf{b-g,} Sequence of measurements (for improved visibility, we show here the raw data after subtracting a slow-varying background) that demonstrates the third-hole occupancy of Q3m.
		In \textbf{(g)}, the square identifies the (Q3m, Q2b, Q2t) (3,1,1) charge state. The gray dashed lines and the labels identify the first three addition lines in Q3m, which partially overlap with the addition lines of Q3b.
	}
	\label{fig:odd_P2P3_1}
\end{figure*}

\begin{figure*}[htbp]
	\centering
	\includegraphics{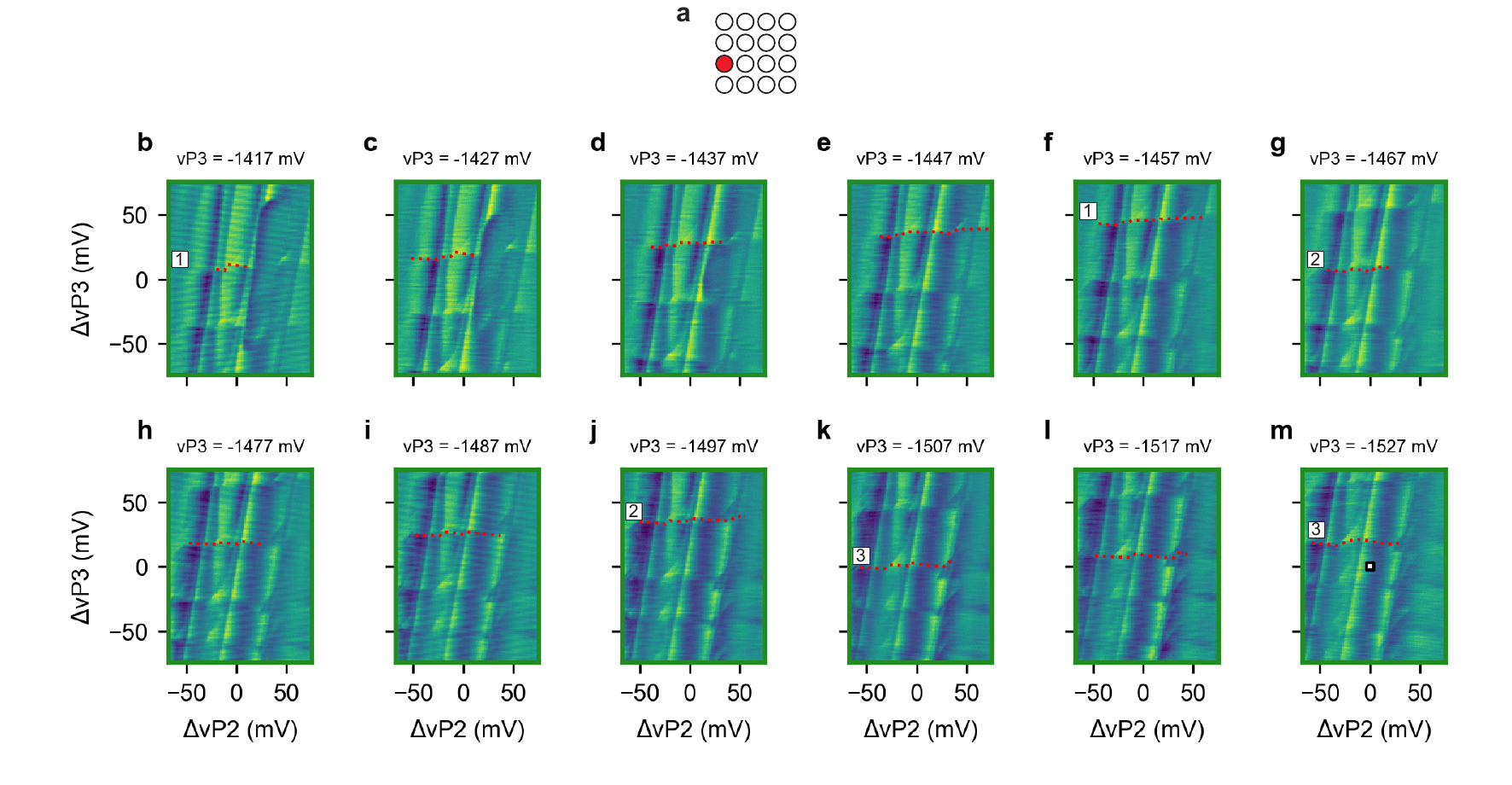}
	\caption{\textbf{Counting the number of holes in Q3b.} 
		\textbf{a,} Schematic of the quantum dot grids labelling the dot Q3b.	
		\textbf{b-g,} Sequence of measurements (for improved visibility, we show here the raw data after subtracting a slow-varying background) that demonstrates the third-hole occupancy of Q3b.
		In \textbf{(g)}, the square identifies the (Q3b, Q2b, Q2t) (3,1,1) charge state. The gray dashed lines and the labels identify the first three addition lines in Q3b, which partially overlap with the addition lines of Q3m.
		Quasi-vertical addition lines tunable by vP2 are associated with a stray dot under P2 outside the array.
	}
	\label{fig:odd_P2P3_3}
\end{figure*}

\begin{figure*}[htbp]
	\centering
	\includegraphics{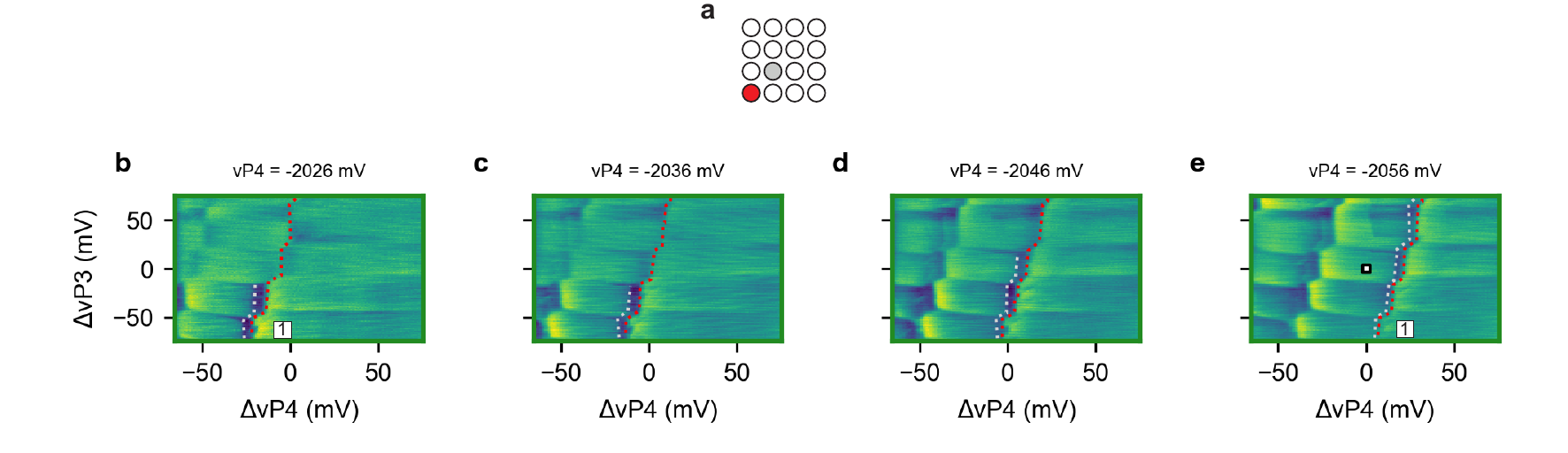}
	\caption{\textbf{Counting the number of holes in Q4b and Q4mb.} 
		\textbf{a,} Schematic of the quantum dot grids labelling the dots Q4b and Q4mb.	
		\textbf{b-e,} Sequence of measurements (for improved visibility, we show here the raw data after subtracting a slow-varying background) that demonstrates the single-hole occupancy of Q4b and Q4mb.
		In \textbf{(e)}, the square identifies the (Q3b, Q4b, Q4mb) (3,1,1) charge state. The dashed lines and the labels identify the first addition lines of Q4b and Q4mb, which mostly overlap. 
		Quasi-horizontal addition lines tunable by vP3 are associated with the quantum dot Q3b.
	}
	\label{fig:odd_P4P3_3}
\end{figure*}

\begin{figure*}[htbp]
	\centering
	\includegraphics{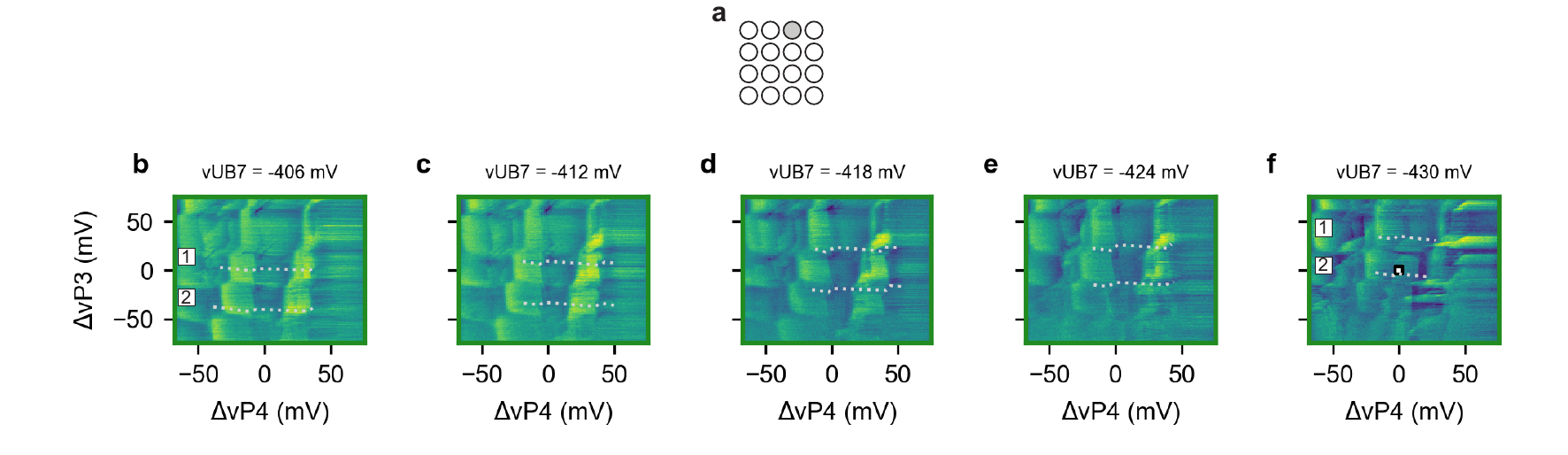}
	\caption{\textbf{Counting the number of holes in Q3t.} 
		\textbf{a,} Schematic of the quantum dot grids labelling the dots Q3t.	
		\textbf{b-f,} Sequence of measurements (for improved visibility, we show here the sum of the signals measured at the SW and SE charge sensor, after removal of a background) that demonstrates the single-hole occupancy of Q3t.
		The addition lines of Q3t are hard to sense due to a nearby stray dot, therefore we rely on following the charge interdot with the more-visible Q4t.
		Differently from other sequences, here we empty the dot Q3t by stepping the voltage on the nearby barrier vUB7.
	}
	\label{fig:odd_P4P3_34}
\end{figure*}

\begin{figure*}[htbp]
	\centering
	\includegraphics{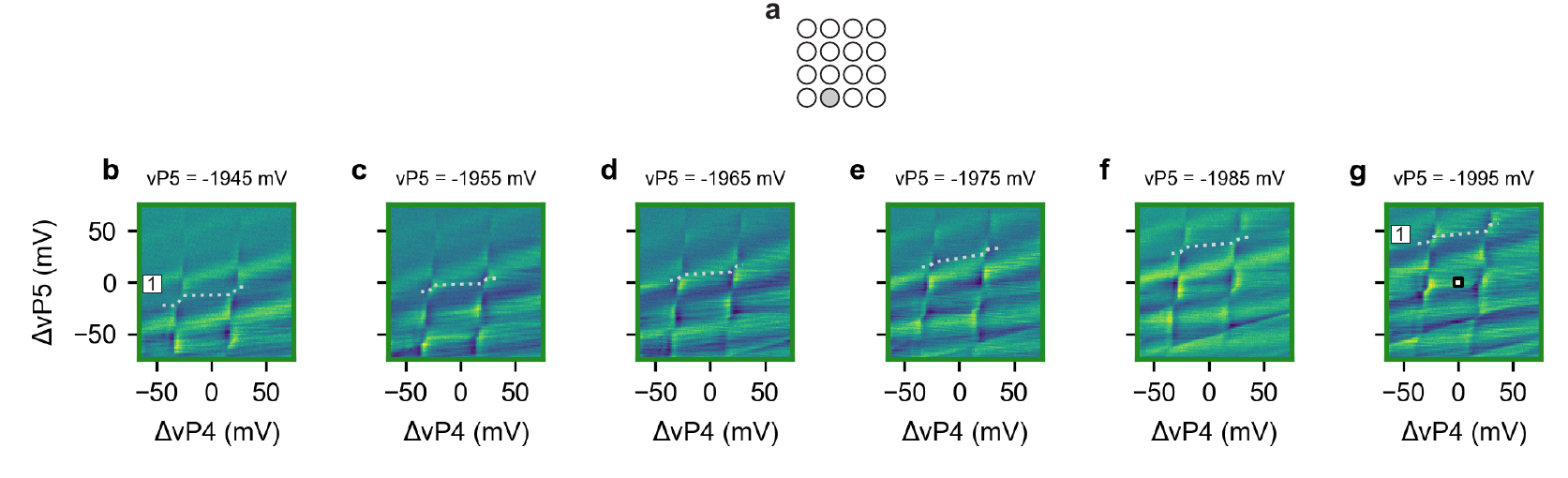}
	\caption{\textbf{Counting the number of holes in Q5b.} 
		\textbf{a,} Schematic of the quantum dot grids labelling the dots Q5b.	
		\textbf{b-g,} Sequence of measurements (for improved visibility, we show here  the raw data after subtracting a slow-varying background) that demonstrates the single-hole occupancy of Q5b.
		In \textbf{(g)}, the square identifies the (Q5b, Q4b, Q4mb) (1,1,1) charge state. The dashed lines and the labels identify the first addition lines of Q5b. Addition lines of Q5b are weakly visible, but can still be tracked by the interdots with Q4b and Q4mb. 
	}
	\label{fig:odd_P4P5_3}
\end{figure*}

\begin{figure*}[htbp]
	\centering
	\includegraphics{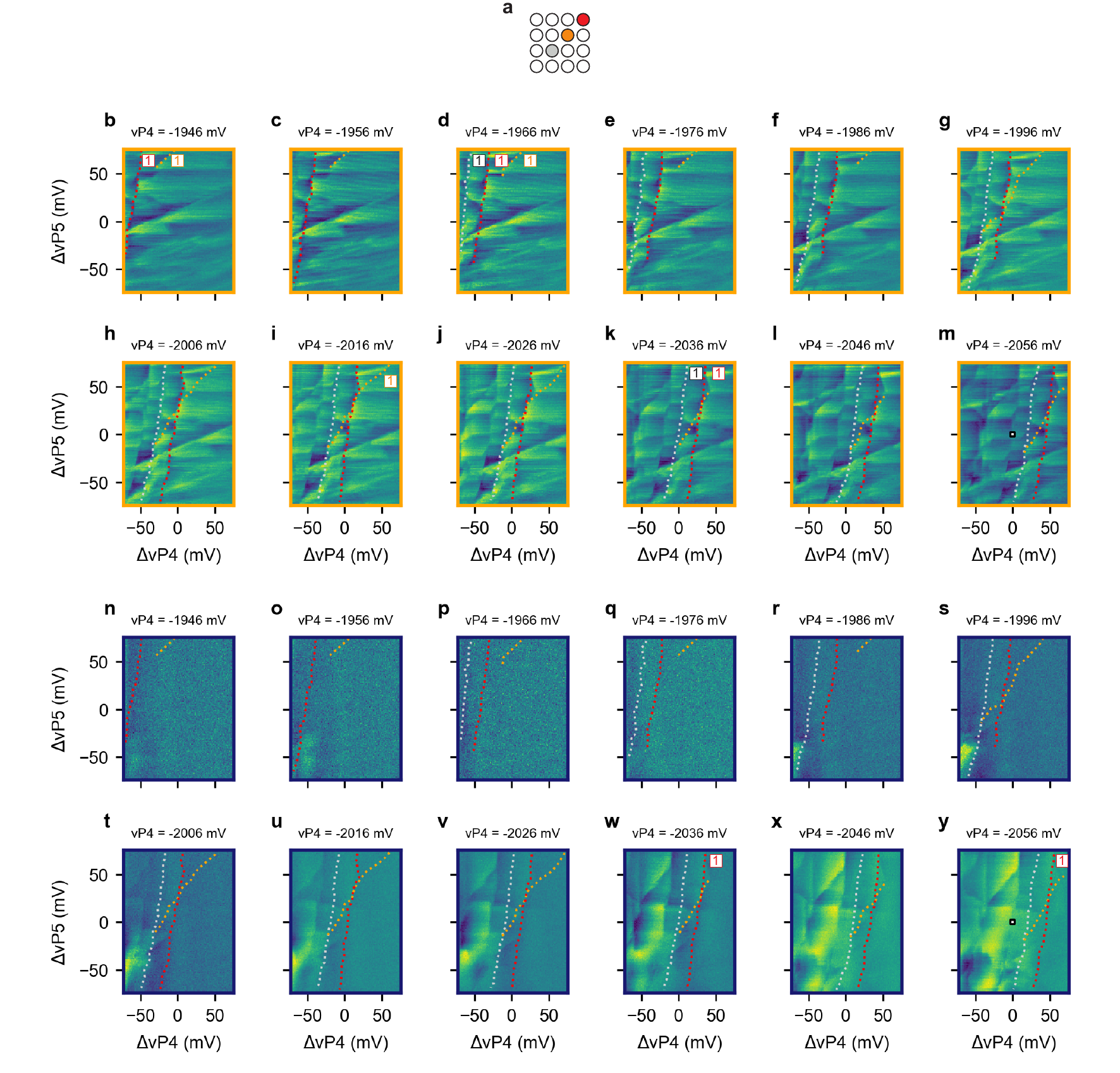}
	\caption{\textbf{Counting the number of holes in Q4t, Q4mt, Q4mb.} 
		\textbf{a,} Schematic of the quantum dot grids labelling the dots Q4t, Q4mt, Q4mb.	
		\textbf{b-m,} and \textbf{n-y,} Sequence of measurements (for improved visibility, we show here the raw data after subtracting a slow-varying background) that demonstrates the single-hole occupancy of the three Q4mb, Q4mt, Q4t dots, collected at the SE and NE charge sensors, respectively.
		The identification of the transition lines is complicated by the slow loading mechanism in the dot Q4mt which mostly appears as a quasi-diagonal line (orange dashed lines). In \textbf{(m)} and \textbf{(y)}, Q5t appears close to the 1-2 charge transition at (0,0), rather than in the 1st hole regime, as shown in the main text Fig.~3c. Such difference is due to an intentional and specific shift in gate voltage of the NE sensor to improve the signal, which results in a modification of the electrostatics of Q4t.
	}
	\label{fig:odd_P4P5_4}
\end{figure*}

\begin{figure*}[htbp]
	\centering
	\includegraphics{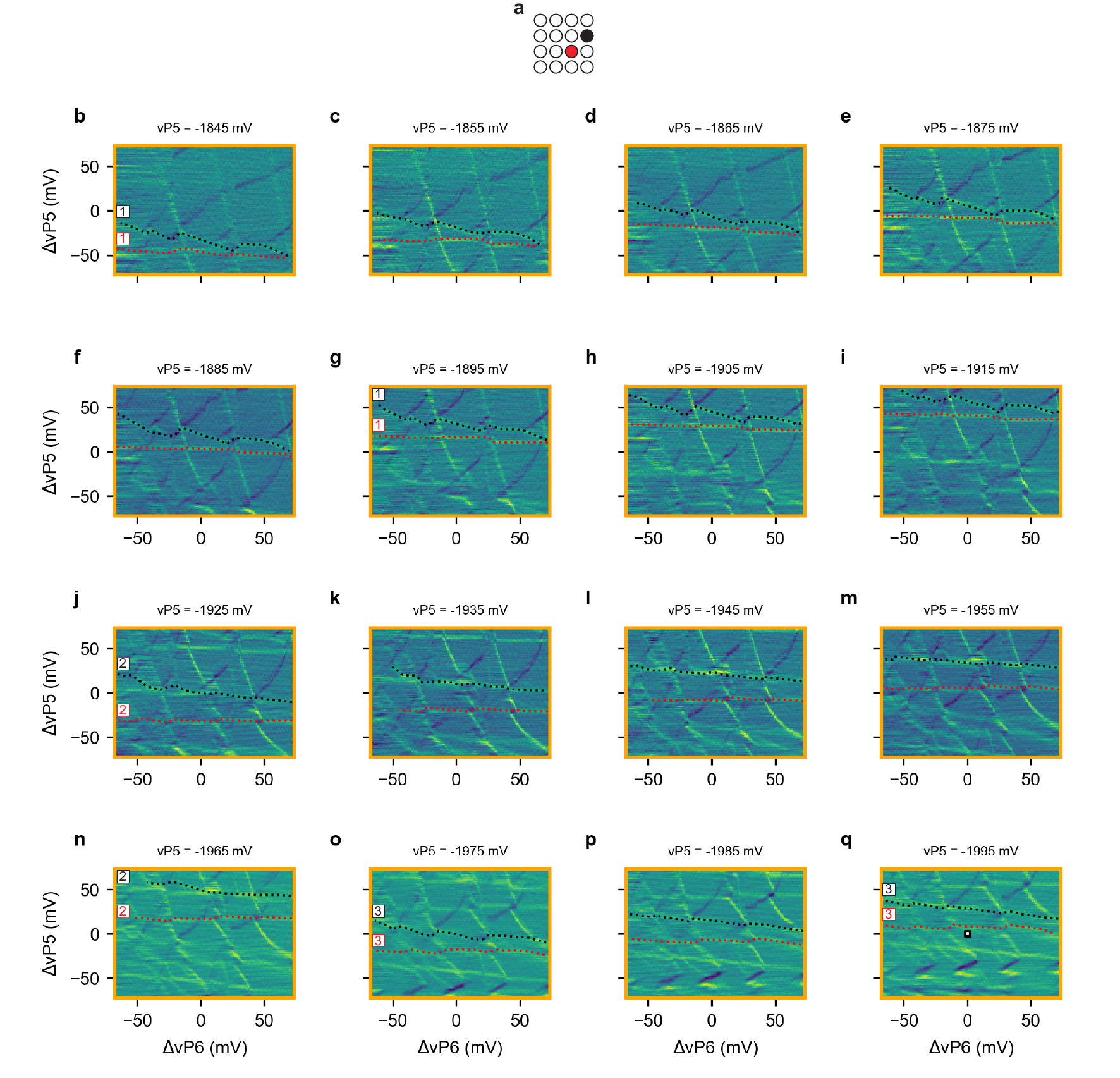}
	\caption{\textbf{Counting the number of holes in Q5m and Q5t.} 
		\textbf{a,} Schematic of the quantum dot grids labelling the dots Q5m and Q5t.	
		\textbf{b-q,} Sequence of measurements (for improved visibility, we show here the partial derivative of the raw data along vP5) that demonstrates the third-hole occupancy of Q5m and Q5t.
		The dashed lines and the labels identify the first three addition lines in Q5m and Q5t.
	}
	\label{fig:odd_P6P5_4}
\end{figure*}

\begin{figure*}[htbp]
	\centering
	\includegraphics{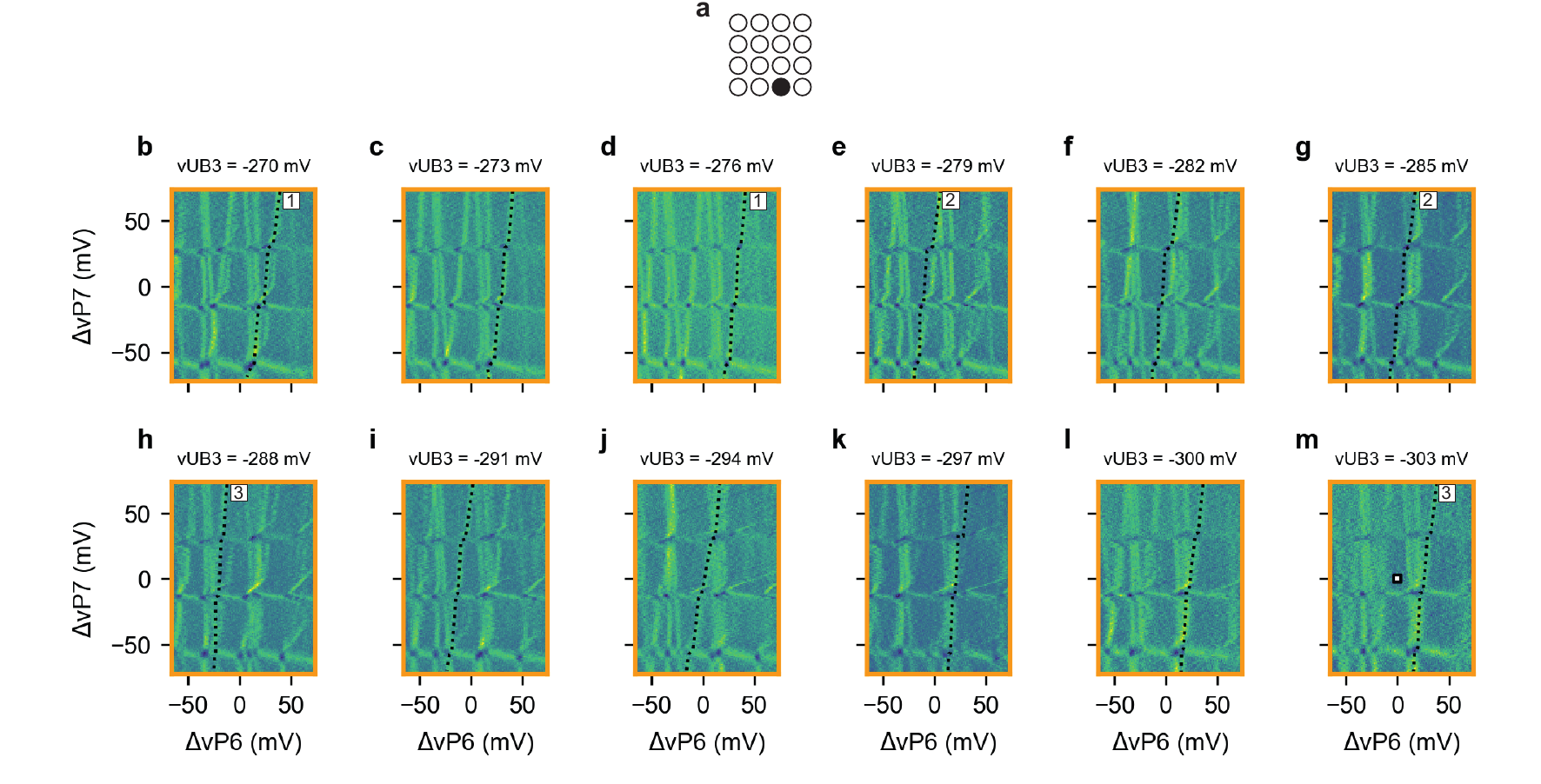}
	\caption{\textbf{Counting the number of holes in Q6b.} 
		\textbf{a,} Schematic of the quantum dot grids labelling the dot Q6b.	
		\textbf{b-m,} Sequence of measurements (for improved visibility, we show here the partial derivative of the raw data along vP6) that demonstrates the third-hole occupancy of Q6b.
		The dashed lines and the labels identify the first three addition lines in Q6b. Due to the presence of a stray dot outside the array tunable by vP6, we step the barrier vUB3 rather than vP6 to isolate the shift of the Q6b addition line.
		In \textbf{(m)}, the square identifies the (Q6b, Q6t, Q7) (3,1,1) charge state.  
	}
	\label{fig:odd_P6P7_4_P6b}
\end{figure*}

\begin{figure*}[htbp]
	\centering
	\includegraphics{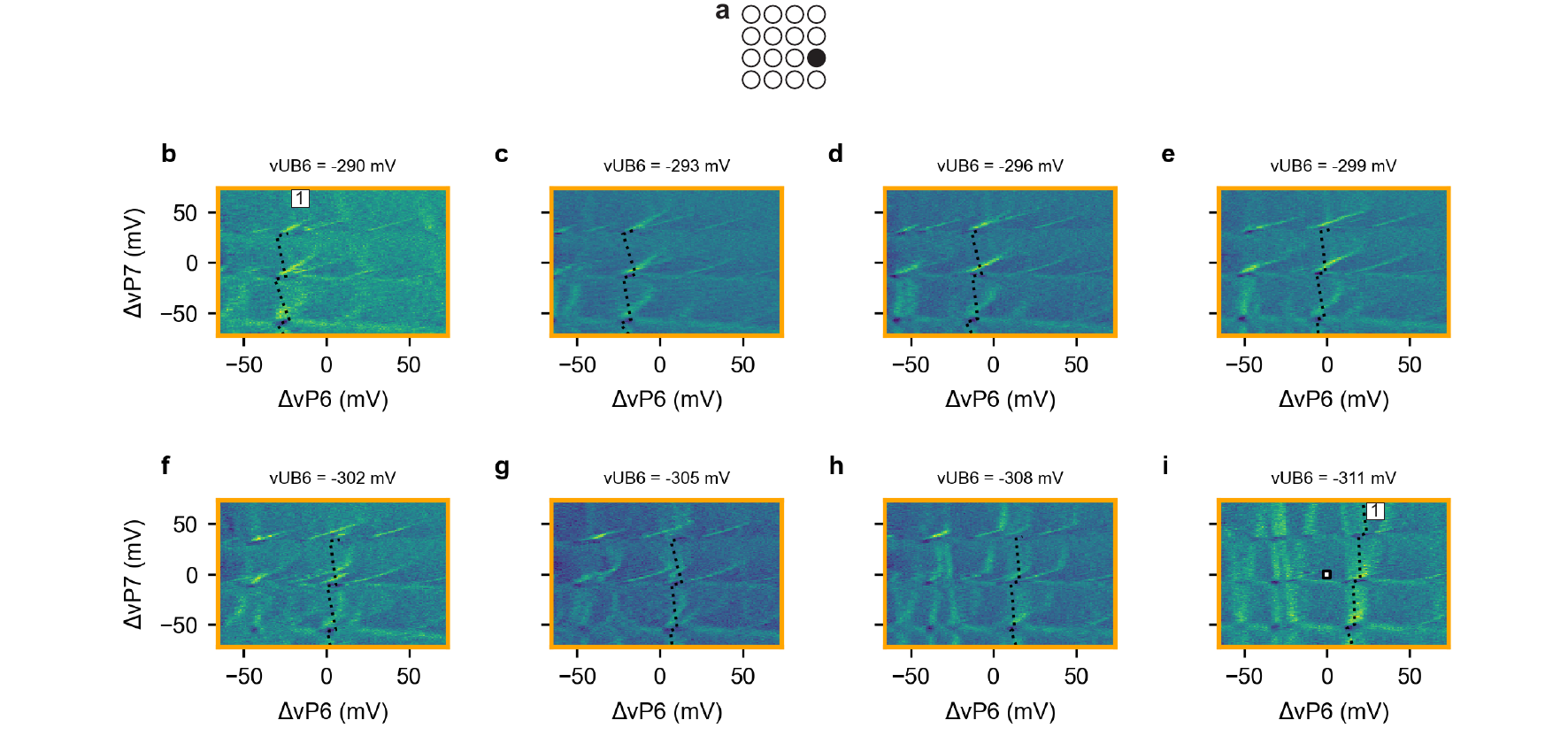}
	\caption{\textbf{Counting the number of holes in Q6t.} 
		\textbf{a,} Schematic of the quantum dot grids labelling the dot Q6t.	
		\textbf{b-i,} Sequence of measurements (for improved visibility, we show here the partial derivative of the raw data along vP6) that demonstrates the single-hole occupancy of Q6t.
		Due to the presence of a stray dot outside the array tunable by vP6, we step the barrier vUB6 rather than vP6 to isolate the shift of the Q6t addition line.
		The dashed lines and the labels identify approximately the first addition line in Q6t. For vUB6 values bigger than $\sim $ -305 mV, the loading in Q6t is slow therefore the addition line latches and is hardly visible
		Nevertheless, in the slow-loading regime, we can still track the interdots with dot Q7.  
		In \textbf{(i)}, the square identifies the (Q6b, Q6t, Q7) (3,1,1) charge state.  
	}
	\label{fig:odd_P6P7_4_P6t}
\end{figure*}

\begin{figure*}[htbp]
	\centering
	\includegraphics{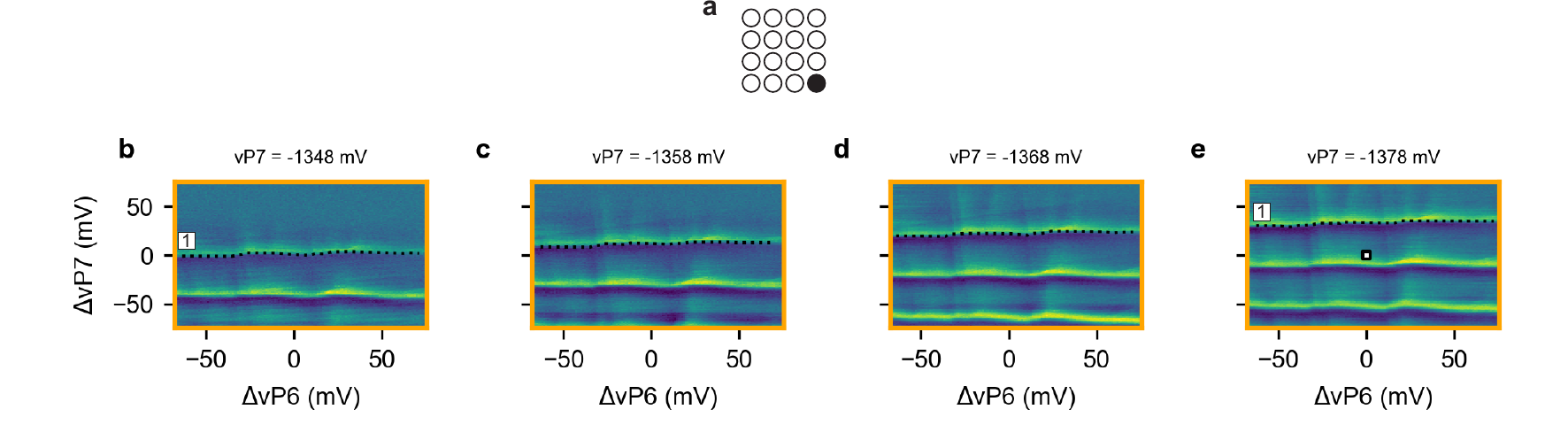}
	\caption{\textbf{Counting the number of holes in Q7.} 
		\textbf{a,} Schematic of the quantum dot grids labelling the dot Q7.	
		\textbf{b-e,} Sequence of measurements (for improved visibility, we show here the raw data after subtracting a background) that demonstrates the single-hole occupancy of Q7.
		In \textbf{(e)}, the square identifies the (Q6b, Q6t, Q7) (3,1,1) charge state. The dashed lines and the labels identify the first addition lines of Q7. 
	}
	\label{fig:odd_P6P7_4}
\end{figure*}

Here, we present sequences of measurements to demonstrate the claim of an odd number of holes confined in each quantum dot. 
To monitor the occupancy of a dot Q$i$, we consider charge stability diagrams of the type vP$i$ versus vP$j$ with $i$, $j \in [1, 7]$ and $i \neq j$.
Because the maximum amplitude of the AWG ramps at the device is only $\sim 200$ mV due to the attenuation in the lines, a single two-dimensional scan is not enough to evaluate directly the number of holes in a quantum dot.
Rather we use a different approach.
We start from the gate voltage regime presented in the main text Fig.~3 and proceed by moving the dc voltage vP$i$ in steps of 10 mV toward more positive voltages until Q$i$ is fully depleted. 
At every step, we take a fast two-dimensional scan that allows to label and count the number of transition lines visible in the available gate window. 
In Suppl. Figs.~\ref{fig:odd_P2P1_1}-\ref{fig:odd_P6P7_4}, we present these measurements starting from an empty dot (left side) and finishing in the original configuration (right side), with a square labelling the (0,0) point in the map signalling the odd occupancy regime.
In analogy to the main text, the red, blue, green and yellow frames of the plots indicate that the data have been measured with the NW, NE, SW and SE charge sensor, respectively.

\clearpage
\bibliography{bibliography}